\documentclass[pra,
,tightenlines
,showkeys
,showpacs
,twocolumn
]
{revtex4-2}

\usepackage{amsmath,amssymb}
\usepackage{bm}
\usepackage[dvipdfmx]{graphicx}
\usepackage{ascmac} 
\usepackage{fancybox}
\usepackage{float}
\usepackage{enumerate}
\usepackage{color}
\usepackage{amsthm}
\usepackage{mathrsfs}
\usepackage{comment}
\usepackage{ulem}
\allowdisplaybreaks[1]
\usepackage{comment}
\newtheorem*{th.}{Theorem}

\newcommand{\tr}{{\rm Tr}}

\newcommand{\1}{{\rm I}}
\newcommand{\2}{{\rm II}}
\newcommand{\3}{{\rm III}}

\begin{document}
	\title{Artificial Relaxation in NMR Experiment}
	\author{Shingo Kukita$^{1)}$}
	\email{kukita@nda.ac.jp} 
	\author{Haruki Kiya$^{2)}$}
	\email{kiya.haruki@kindai.ac.jp }
	\author{Yasushi Kondo$^{2)}$}
	\email{ykondo@kindai.ac.jp}
	\affiliation{$^{1)}$Department of Computer Science, National Defence Academy of Japan, 
		1-10-20, Hashirimizu, Yokosuka, Japan}
	\affiliation{$^{2)}$Department of Physics, Kindai University, Higashi-Osaka 577-8502, Japan}
	
\begin{abstract}
	
Environmental noises cause the relaxation of quantum systems and decrease the precision of operations.
Apprehending the relaxation mechanism via environmental noises is essential for building 
quantum technologies. Relaxations can be considered a process of information dissipation 
from the system into an environment with infinite degrees of freedom (DoF).  
According to this idea, a model of artificial relaxation has been proposed and demonstrated in NMR experiments.  
Although this model successfully understood the central idea of relaxation, 
we observed recursive behavior, which is non-ideal to describe relaxation, because of few DoF of 
the ``artificial environment''. In this paper, we extend the approach of the artificial environment and discuss, theoretically and experimentally,
how many DoF of the environment are necessary for realizing ideal relaxation behavior. 
Our approach will help us thoroughly understand the concept of relaxation.
\end{abstract}
\maketitle

\section{introduction}

A quantum system suffering from environmental noises is an open system \cite{Weiss}.
Realistic quantum devices, which are utilized in emerging technology such as quantum sensing~\cite{helstrom1976quantum,caves1981quantum,holevo2011probabilistic}, communication~\cite{ekert1991quantum,gisin2007quantum,chen2021integrated} and computation~\cite{bennett2000quantum,Nielsen2000,nakahara2008quantum}, are regarded as open systems.
To improve quantum technologies, we desire to control the behaviors of such systems at will.
Microscopic picture of open systems has also been studied in the context of thermodynamics\cite{PhysRev.129.1880,PhysRevE.50.888,Zurek2007,PhysRevE.79.061103,PhysRevLett.106.040401,PhysRevLett.119.100601}, which is of considerable interest since Maxwell's era.n the thermodynamics of a system~\cite{PhysRevE.81.051135,PhysRevLett.115.120403}.
Therefore, it is necessary from both viewpoints of science and technology to investigate 
behaviors of open systems.
To deeply apprehend open systems, many efforts have been carried out, for example, utilizing optics~\cite{liu2011experimental}, ultracold atoms~\cite{RevModPhys.83.863}, trapped ions~\cite{RevModPhys.75.281}, and cold electric circuits~\cite{pekola2015towards}.

An open quantum system is a subsystem of a large system.  
The open system interacts with the other part of the large one, called 
an environment and the total dynamics is governed by unitary evolution.
If we focus only on the subsystem, we witness the leakage of ``information" from it 
to the environment owing to the interaction between them.
As the total dynamics is unitary, this information will return to the subsystem someday.
However, the environment's degrees of freedom (DoF) are uncountable, 
and hence its backflow must only occur once the Universe dies.
This is an interpretation of relaxation in an open system and leads us to the idea of an artificial open system by which we can intuitively 
understand the concept of open systems.

\begin{figure}[tbh]
	\begin{center}
		\includegraphics[width=60mm]{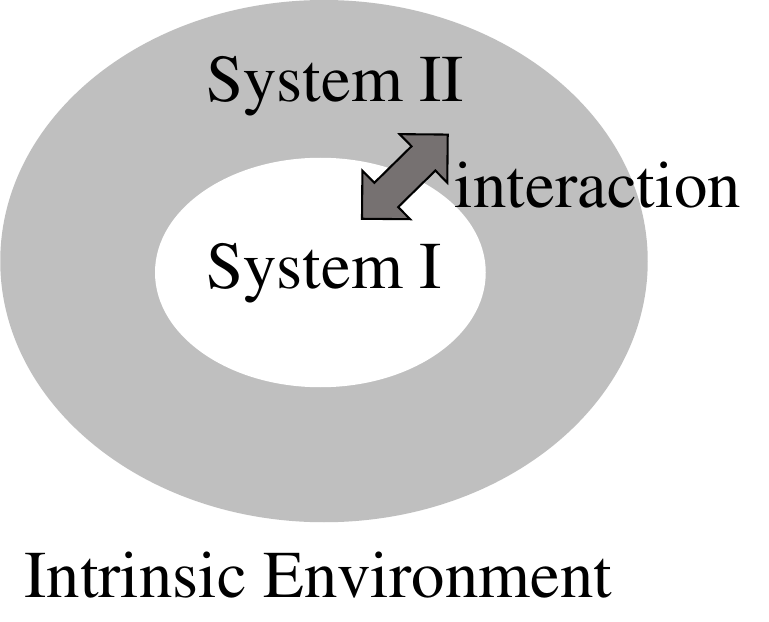}
		\caption{The concept of the artificial open system. 
			System~I is surrounded by  System~II.
			Both systems weakly interact with an intrinsic environment.
			Due to their interaction, System~I exhibits a relaxation-like behavior in a particular time scale.
			\label{fig:idea}
		}
	\end{center}
\end{figure}	

\begin{figure}[H]
\includegraphics[width=35mm]{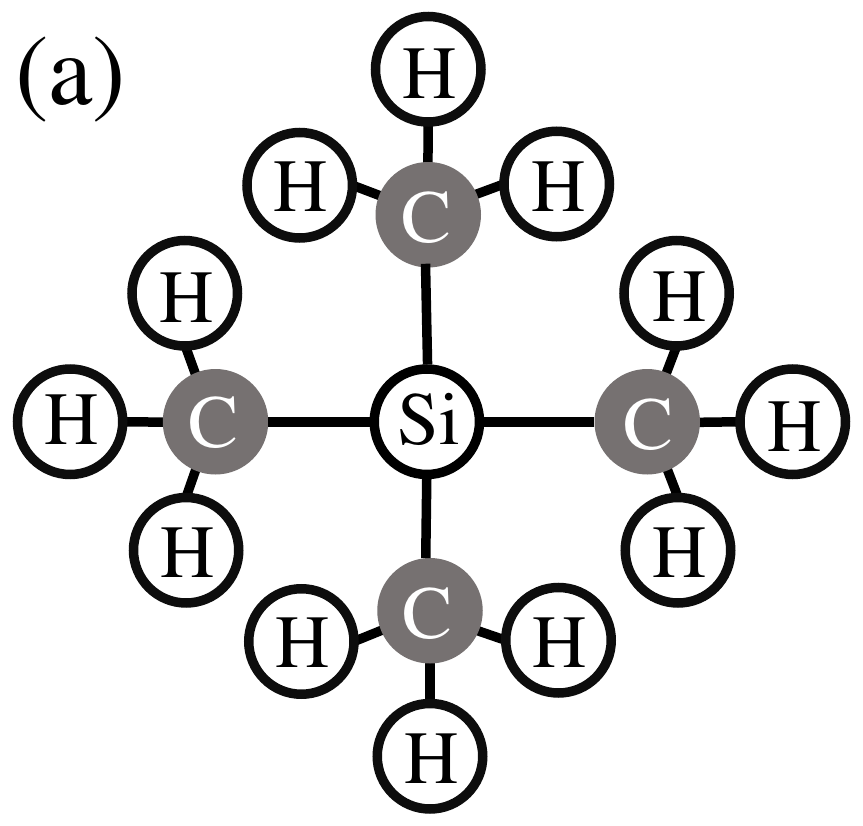}
\includegraphics[width=45mm]{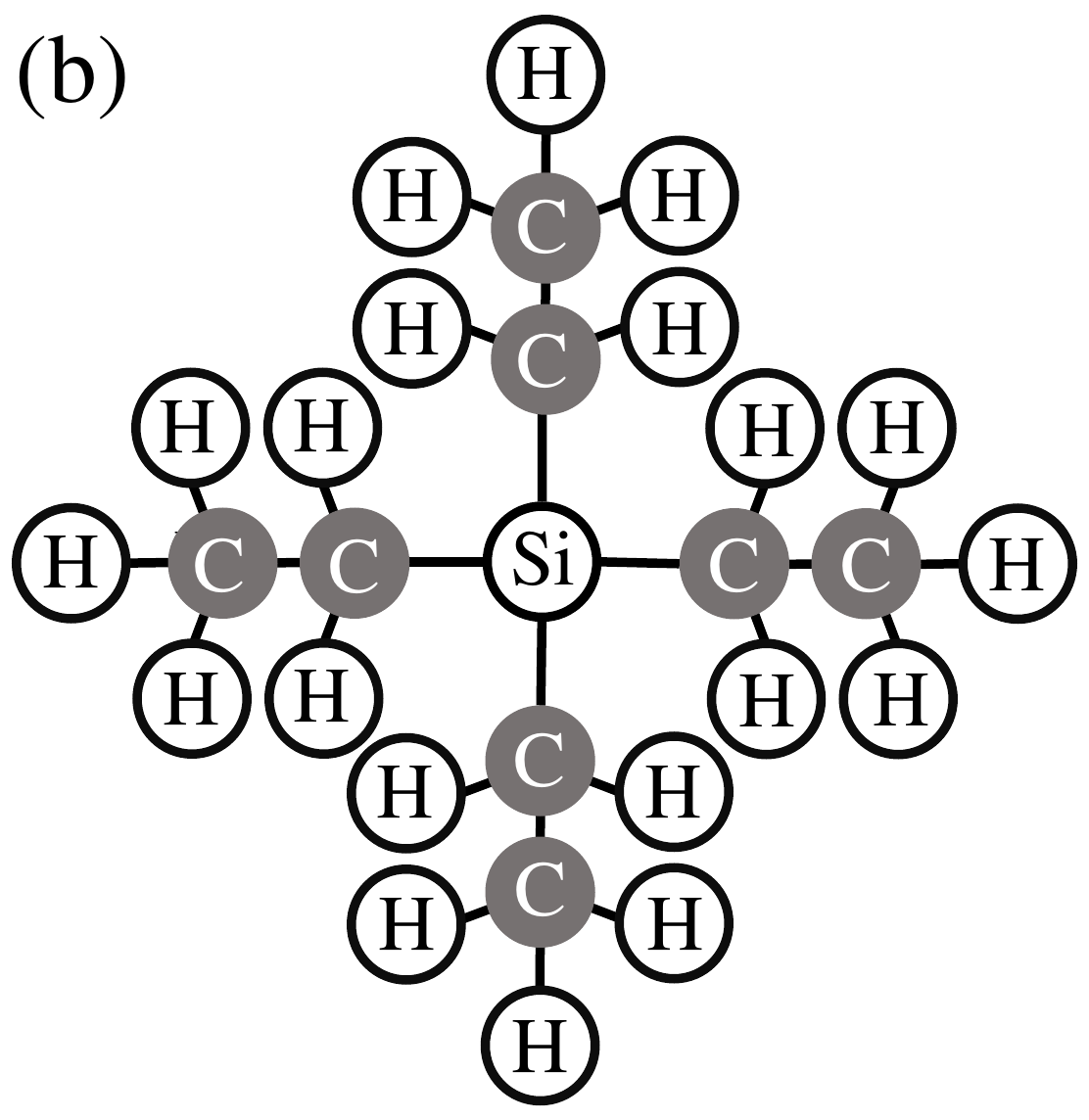}
\caption{
	Molecules to be considered.
	(a) Tetramethylsilane (TMS, Si(CH$_{3}$)$_{4}$).
	(b) Tetraethylsilane (TES, Si(C$_{2}$H$_{5}$)$_{4}$)). 
	In both panels, the central nucleus of Si works as System~I, and the surrounding H nuclei are System~II. 
	The nuclei of C do not affect the FID signal of Si because they have no spin.}
\label{fig:TMS}
\end{figure}

A theoretical model of an artificial open system is schematically represented as follows~\cite{PhysRevA.96.032303,kondo2018,Ho_2019,Kukita_2020}.
The model comprises System~I and II (Fig.~\ref{fig:idea}), 
both of which are well-controllable and weakly interact with an intrinsic environment.
System~I, whose dynamics we focus on, interacts with System~II, representing the ``artificial environment''.
In a particular time scale, System~I exhibits relaxation-like behavior due to the interaction with System~II, although the information, which went away from System~I by this behavior, 
will come back in a realistic time scale; controllability of System~II usually implies few DoF.
More DoF of System~II will make a more plausible relaxation of System~I.
Reference~\cite{PhysRevA.96.032303,Ho_2019} provided an explicit form of this model 
using multiple qubits (two-level systems).
System~I is represented as a qubit, while System~II consists of the other qubits that are coupled 
to the qubit of System~I.
Presuming a simple interaction between them and a particular initial state, 
we can explicitly solve this model.
The expectation value $\langle\sigma_x \rangle$ ($\sigma_{x}$ is the $x$ component of the Pauli matrices) 
of System~I exhibits relaxation-like behavior due to its interaction with System~II, 
while we also witness its recursion as expected.
This model was implemented in an NMR experiment using molecules of Tetramethylsilane 
(TMS, Fig.~\ref{fig:TMS}(a)). We can directly observe $\langle \sigma_x \rangle$ 
as a Free Induction Decay (FID) signal.

In this paper, we consider a generalization of this model 
by adding more DoF to System~II than the previous one.
The expectation value $\langle \sigma_x \rangle $ of System~I also 
shows a relaxation-like behavior. 
However, the recursion is suppressed thanks to more DoF of System~II
while we can still analytically solve the dynamics. 
The dynamics of this model mimic environmental noises more plausibly 
than that of the previous one.
A part of the corresponding experiments is implemented using molecules of Tetraethylsilane (TES, Fig.~\ref{fig:TMS}(b)).

The remainder of this paper is organized as follows.
Section~\ref{sec:TMS} reviews the results in Refs.~\cite{Ho_2019} and \cite{Kukita_2020}.
We introduce in Sec.~\ref{sec:TES} an extended proof-of-principle model of an open system 
and propose the corresponding NMR experiment using TES.
Section~\ref{sec:conclusion} is devoted to summarizing this work.

\section{review of ``relaxation'' in Tetramethylsilane}
\label{sec:TMS}

A theoretical model of an artificial open system and the corresponding NMR experiment have been proposed \cite{PhysRevA.96.032303,Ho_2019}.
The model is comprised of System~I and II (Fig.~\ref{fig:idea}).
System~I is a qubit, while System~II is a composite system of multiple qubits.
System~I is regarded as the target whose dynamics we focus on.
On the other hand, System~II represents the ``artificial environment''.
The qubits in System~II are coupled to System~I via a simple interaction, but there is no interaction among them.
The dynamics of System~I is affected by ``noise" due to this interaction. 
Despite the relatively many DoF in the total system, this dynamics can be solved analytically
when we only focus on the phase dynamics or the expectation value $\langle\sigma_{x}\rangle$ of System~I.
Reference~\cite{Kukita_2020} refines the theory in a more sophisticated way and 
sheds light on the extensibility of this model. An NMR experiment corresponding to the model is performed with molecules of TMS. 
A TMS molecule is comprised of Si, C, and H (Fig.~\ref{fig:TMS}(a)).
In the experiment, the Si nucleus is System~I, while the surrounding H nuclei represent System~II.
The C nuclei can be ignored because 99~\% of them are ${}^{12}$C and have no spin. 
The expectation value $\langle \sigma_x \rangle$ of System~I in theory can experimentally be observed 
as a free induction decay (FID) signal of Si. 
As reviewed below, we see that this signal exhibits a relaxation-like behavior 
due to the interaction with H nuclei.
The results of the theoretical model and the experiments have an excellent agreement
when we take into account unavoidable real relaxation. 

We introduce the mathematical description of FID dynamics of TMS and corresponding experiments.
Consider a composite system consisting of $(N+1)$ qubits.
The $0$th qubit represents System~I while the other $N$ qubits $(k=1\sim N)$ do System~II.
The following Hamiltonian governs the dynamics:
\begin{align}
	H&=H_{0}+\sum^{N}_{k=1}H^{(k)}_{J},
	\nonumber \\
	H_{0}
	&:=\sum^{N}_{i=0}\frac{\omega_{i}}{2}\sigma^{(i)}_{z},~~H^{(k)}_{J} 
	:=\frac{J}{4}\sum_{\mu=x,y,z}\sigma^{(0)}_{\mu}\sigma^{(k)}_{\mu},
\end{align}
where $\sigma^{(i)}_{\mu}$($i=0,\cdots,N$, $\mu=x,y,z$) are the Pauli matrices 
acting on the relevant qubit; for instance, $\sigma^{(0)}_{z}$ is 
$\sigma_{z}\otimes \sigma_{0}\otimes\cdots \sigma_{0}$.
The parameters  $\omega_{i}~(i=0,1,\cdots,N)$ represent the resonance frequencies 
of the corresponding qubits while $J$ is the coupling strength 
between the $0$th qubit and the qubits in System~II.
Note that all the qubits in System~II are coupled to the qubit of System~I with the same interaction strength.
Actually, even if $J$ has $k$-dependence, the dynamics discussed below can be solved analytically 
(Appendix~\ref{sec:app_a}).
We assume that there is no interaction among qubits in System~II.
Changing from the laboratory frame to the rotating one with respect to $H_{0}$ provides
\begin{equation}
\tilde{H}\approx \frac{J}{4}\sum^{N}_{k=1}\sigma^{(0)}_{z}\sigma^{(k)}_{z},
\label{eq:hamiltonian0}
\end{equation}
where we use the rotating wave approximation (RWA) with the assumption of $|\omega_{0}-\omega_{k}|\gg J~(\forall k=1\sim N)$.

We consider the following initial state at $t=0$:
\begin{align}
	\rho(0)
	=&\frac{1}{2^{N+1}}
		\underbrace{\left(\sigma^{(0)}_{0}+\sigma^{(0)}_{+}+\sigma^{(0)}_{-}\right)}_{\rm System~I}
		\underbrace{\prod^{N}_{k=1}\sigma^{(k)}_{0}}_{\rm System~II}
		\nonumber \\
		&=\frac{1}{2^{N+1}}\left(\sigma^{(0)}_{0}+\sigma^{(0)}_{+}+\sigma^{(0)}_{-}\right),
	\label{eq:initial0}
\end{align}
in the rotating frame, where $\sigma_0$ denotes the $2 \times 2$ identity matrix and 
$\sigma_\pm = (\sigma_x \pm i \sigma_y)/2$. 
Note that $\sigma^{(i)}_{0}$ for any $i=0,\cdots,N$ is just the $2^{N}\times 2^{N}$ identity matrix.
The following Liouville-von Neumann equation describes the dynamics:
\begin{equation}
\frac{d \rho(t)}{d t}=-i[\tilde{H},\rho(t)],
\label{eq:hamiltonian}
\end{equation}
where $[\bullet,\bullet]$ represents the commutator of two operators.
The solution of this equation is
\begin{align}
\rho(t) &=
\frac{\sigma^{(0)}_{0}}{2^{N+1}}
+\frac{\sigma^{0}_{+}}{2}\prod^{N}_{k=1}A^{(k)}(t)
+\frac{\sigma^{(0)}_{-}}{2}\prod^{N}_{k=1}A^{(k)}(t),
\nonumber \\
A^{(k)}(t)&=\cos(J t/2)\frac{\sigma^{(k)}_{0}}{2}+\sin(J t/2)\frac{\sigma^{(k)}_{z}}{2}.
\label{eq:solution0}
\end{align}
See Appendix \ref{sec:app_a} for the details.

In the case of the TMS molecule, the central Si nucleus corresponds to System~I 
while the twelve H nuclei are System~II, which means $N=12$.
The state of this composite system after applying a $\pi/2$-pulse to Si is 
described by Eq.~(\ref{eq:initial0}) up to normalization~\cite{kondo_2016}.
The parameter $J$ is determined via experiments~\cite{Ho_2019}.
The expectation value $\langle \sigma_x \rangle$ of System~I is described by
\begin{align}
	S^{\rm TMS}(t):=&\tr\left(\sigma^{(0)}_{x}\rho(t)\right)=\cos^{N}(J t/2),
	\label{eq:signalTMS0}
\end{align}
which corresponds to the FID signal of Si in the experiment.
Note that $\tr \left(\sigma_{y}^{(0)}\rho(t)\right)=0$, which should be taken to be $0$ also in the experiment below.

\begin{figure}[h]
	\begin{tabular}{cc}
	\begin{minipage}[t]{0.9\hsize}
	\centering
	\includegraphics[width=80mm]{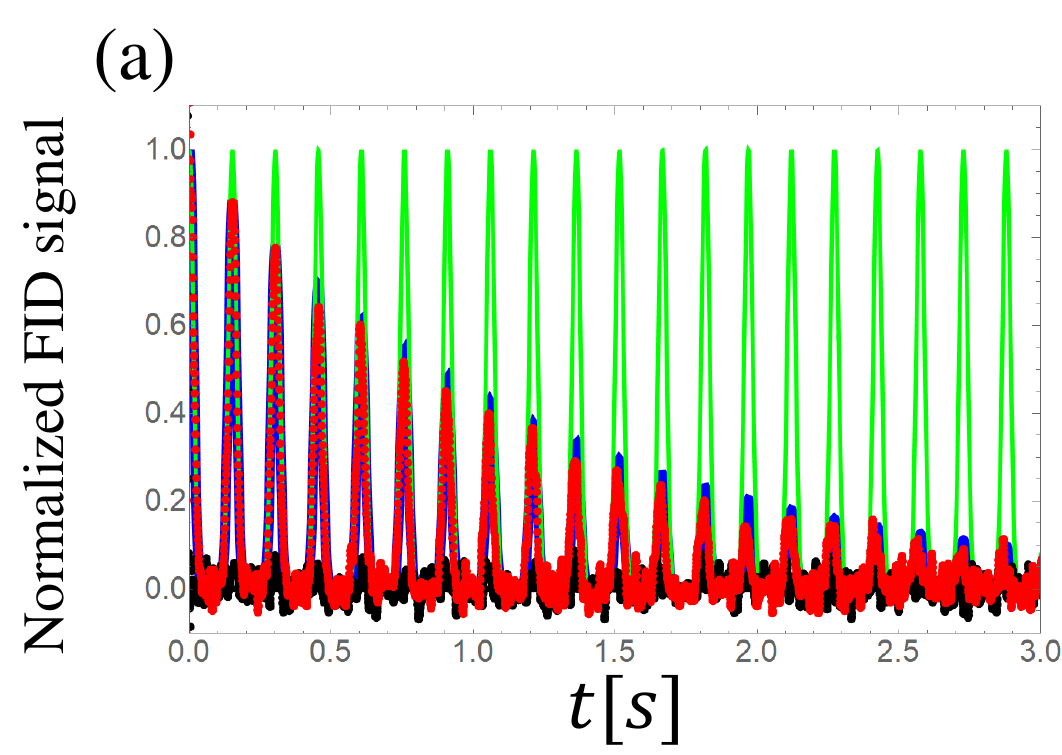}
	\end{minipage} \\
	\begin{minipage}[t]{0.9\hsize}
	\centering
	\includegraphics[width=80mm]{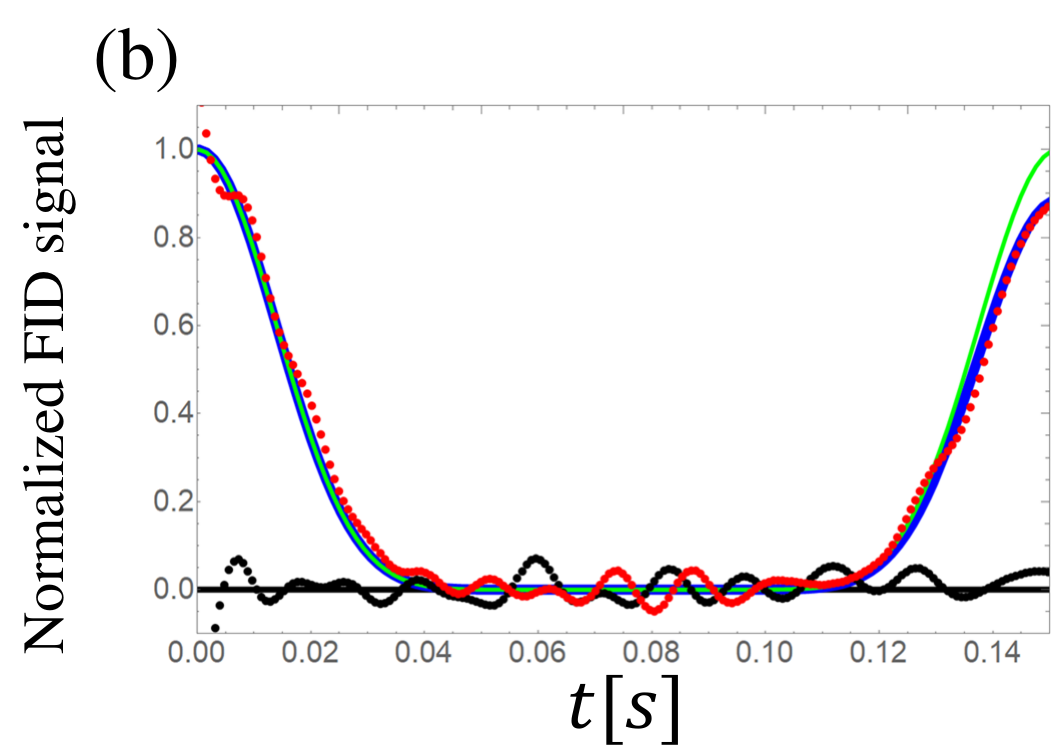}
	\end{minipage}
	\end{tabular}
	\caption{(color online) The dynamics of $S^{\rm TMS}(t)$, $S^{\rm TMS}_{\cal L}(t)$, and 
	the corresponding experimental result in (a) a long-time region $t\in[0,3.0]$~s, and 
	(b) a short time region $t \in [0,0.15]$~s.
In both panels, the green line represents $S^{\rm TMS}(t)$ while the blue line is $S^{\rm TMS}_{\cal L}(t)$.
The red dots are the experimental result of the (real part of) FID signal of Si.
The black dots are the imaginary part of the FID signal, corresponding to the expectation value of $\sigma_{y}$, and should be taken $0$ in this situation.
We compare $S^{\rm TMS}(t)$, $S^{\rm TMS}_{\cal L}(t)$ and the red dots.}
	\label{fig:TMSwithEXP}
	\end{figure}

Figure~\ref{fig:TMSwithEXP} compares 
$S^{\rm TMS}(t)$ with experimental results.
In the long time region shown in Fig.~\ref{fig:TMSwithEXP}(a), $S^{\rm TMS}(t)$ 
predicts recursive dynamics (the green line)  because of the finite DoF in System~II. 
Note, however, that $S^{\rm TMS}(t)$ well reproduces a relaxation-like behavior in a short time scale, 
as shown in Fig.~\ref{fig:TMSwithEXP}(b). If we stop observing this dynamics 
at 0.11~s, we cannot observe the recursive dynamics and may ``experimentally'' judge 
that System~I has completely relaxed. It is very similar to the case that  
we cannot observe a recursive behavior in real relaxation because of the finite life of the Universe. 
The red dots show the FID signal of Si and agree very well with the green line at $t \in [0, 0.11]$~s. 
It implies that we successfully reproduce relaxation phenomena theoretically and 
experimentally with the NMR technique. 

We can compare $\langle \sigma_x \rangle $ of System~I with an experimental FID signal of Si in the whole time region by considering effects from an {\it intrinsic} environment.
In the case of liquid-state NMR, the origin of such effects is magnetic impurities in the solvent.
They are described by the following GKSL term~\cite{GKS,Lindblad1976} 
with infinite temperature approximation: 
\begin{align}
	{\cal L}[\rho] &:=\sum^{N}_{i=0}{\cal L}^{(i)}[\rho],
	\nonumber \\
	{\cal L}^{(i)}[\rho]
	&:=\frac{\gamma_{i}}{2}\left(\sum_{\mu=\pm}\sigma^{(i)}_{\pm}\rho\sigma^{(i)}_{\mp}-\rho\right).
	\label{eq:lindbladian0}
\end{align}
We assume that $\gamma_{i}$ for all $k=1,\cdots,N$ are identical: $\gamma_{i}=\gamma_{\rm II}$ for $k=1,\cdots,N$ while we let $\gamma_{\rm I}$ denote $\gamma_{0}$.
As in the case of $J$, the $k$ dependence of $\gamma_{\rm II}$ does not affect the analytical solvability of the model.
We solve the following GKSL equation
instead of Eq.~(\ref{eq:hamiltonian}),
\begin{equation}
	\frac{d \rho(t)}{d t}=-i[\tilde{H},\rho(t)]+{\cal L}[\rho(t)].
\end{equation}
The solution of the above equation still has a simple form (Appendix~\ref{sec:app_a}$ $),
\begin{align}
	\rho_{\cal L}(t)
	&=\frac{\sigma^{(0)}_{0}}{2}
	+e^{-\gamma_{\rm I}t/2}\frac{\sigma^{(0)}_{+}}{2}\prod^{N}_{k=1}A_{\cal L}^{(k)}(t)
	\nonumber \\
	&+e^{-\gamma_{\rm I}t/2}\frac{\sigma^{(0)}_{-}}{2}\prod^{N}_{k=1}A_{\cal L}^{(k)\dagger}(t),
	\nonumber\\
	A_{\cal L}^{(k)}(t)
	&=e^{-\gamma_{\rm II} t/2}\bigg(\frac{\gamma_{\rm II}}
	{\Omega}\sin(\Omega t/2)+\cos(\Omega t/2)\bigg)\frac{\sigma^{(k)}_{0}}{2}
	\nonumber \\
	&+e^{-\gamma_{\rm II} t/2}\bigg( \frac{- i J}{\Omega}\sin(\Omega t/2)\bigg)\frac{\sigma^{(k)}_{z}}{2},
	\label{eq:solution0noise}
\end{align}
where $\Omega=\sqrt{J^{2}-\gamma^{2}_{\rm II}}$.
We assume $J>\gamma_{\rm II}$ and thus $\Omega$ is real.
The parameters $(\gamma_{\1},\gamma_{\rm II})= (0.21, 0.1)$~s$^{-1}$ are determined 
via experiments~\cite{Ho_2019}.
The expectation value $\langle \sigma_{x}\rangle$ in this noisy case, $S^{\rm TMS}_{\cal L}(t)$, is described by
\begin{align}
	S^{\rm TMS}_{\cal L}(t)
	&:=\tr\left(\sigma^{(0)}_{x}\rho_{{\cal L}}(t)\right)
	\nonumber \\
	&=e^{-\gamma_{0} t/2}\left(e^{-\gamma_{\rm II} t/2}\bigg(\frac{\gamma_{\rm II}}{\Omega}\sin(\Omega t/2)+\cos(\Omega t/2)\bigg)\right)^{N}, 
	\label{eq:signalTMS}
\end{align}
and is shown with the blue curve in Fig.~\ref{fig:TMSwithEXP}. We found that 
$S^{\rm TMS}_{\cal L}(t)$ can reproduce the FID signal of Si in the whole time region 
up to 3.0~s. It implies that our model can well be realized in NMR experiments 
even with an intrinsic environment.

\section{More Degrees of Freedom of  ``artificial environment''}
\label{sec:TES}

We propose an extension of the above theoretical model  by increasing 
DoF of the artificial environment with experimental realization in mind. For this purpose, 
we add more qubits in the artificial environment like in Fig.~\ref{fig:TMS}(b) 
and call them System~III.
System~II and III interact with System~I in different coupling strengths and are coupled to each other.
Similarly to the model in \S~II, the expectation value 
$\langle \sigma_x \rangle$ of System~I has a relaxation-like behaviour 
owing to its interaction with System~II and III.
The relaxation-like behavior is more plausible than in the previous model, thanks to the more DoF. 

\subsection{Case when Rotating Wave Approximation is valid}  
\label{RWA}
\begin{figure}[h]
	\begin{center}
		\includegraphics[width=80mm]{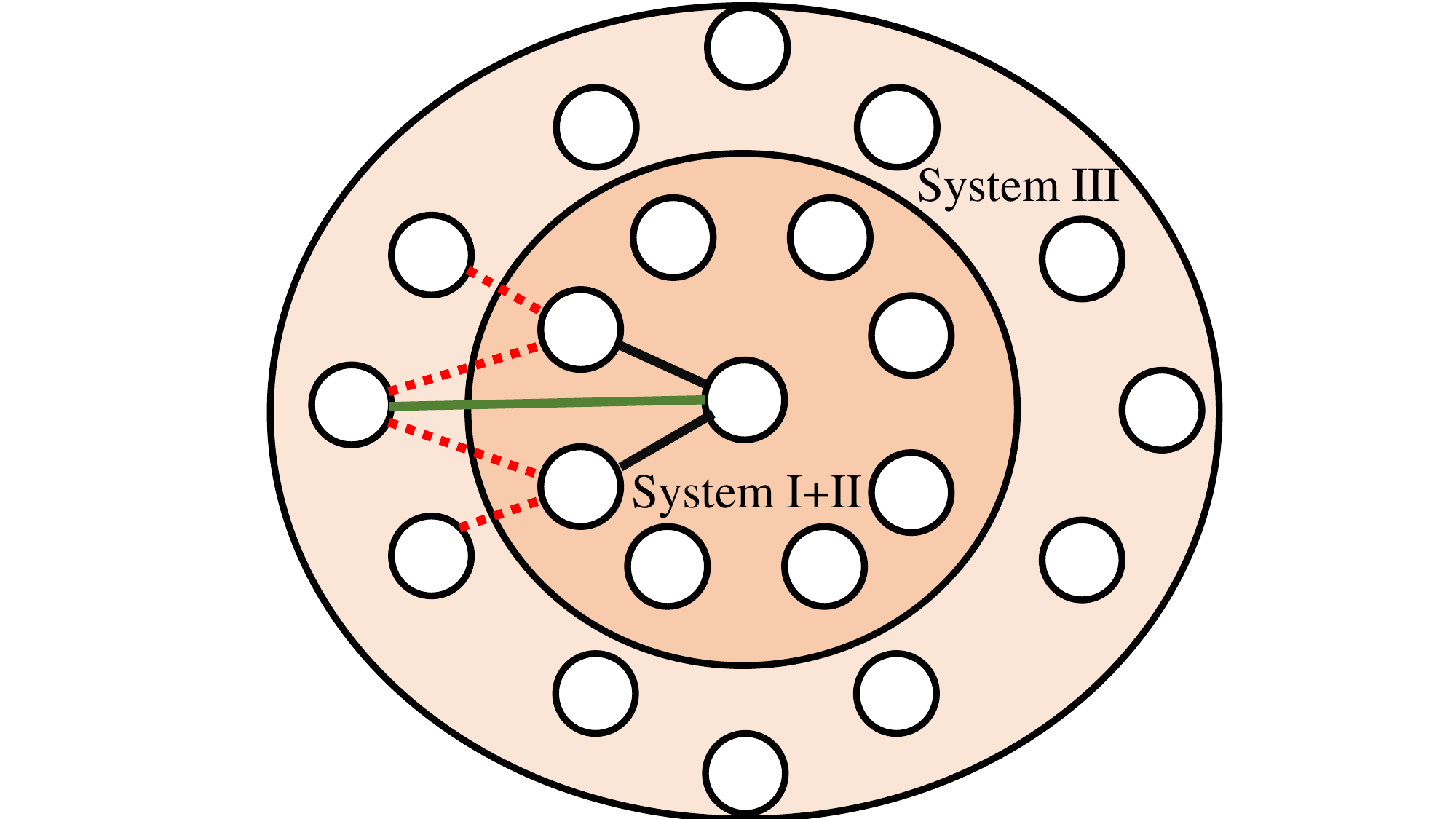}
		\caption{(color online) The schematic picture of the proposed model.
		System~I interacts with System~II (black lines) and System~III (green line).
		There are interactions between System II and III (red dotted lines).
			\label{fig:System III}
		}
	\end{center}
\end{figure}

Consider a multiple-qubit model shown in Fig.~\ref{fig:System III}.
The central $0$th qubit is System~I.
System~II (III) comprises $M$ ($N$) qubits.
System~II and III interact with each other, while there is no interaction 
among qubits in System~II (III).

The Hamiltonian is given as
\begin{align}
H&=H_{0}+H^{\rm in}+H^{\rm ex},
\nonumber \\
H_{0} &=\frac{\omega_{\1}}{2}\sigma_{z}^{(0)}
+\frac{\omega_{\2}}{2}\sum^{M}_{\alpha=1}\sigma^{(\2,\alpha)}_{z}
+\frac{\omega_{\3}}{2}\sum^{N}_{\beta=1}\sigma^{(\3,\beta)}_{z},
\nonumber\\
H^{\rm in}&=\sum^{M,N}_{\alpha,\beta=1}\sum_{\mu=x,y,z}
\frac{J^{(\alpha,\beta)}_{23}}{4}\sigma^{(\2,\alpha)}_{\mu}\sigma^{(\3,\beta)}_{\mu},
\nonumber \\
H^{\rm ex}&=\sum^{M}_{\alpha=1}\left(\sum_{\mu=x,y,z}\frac{J_{12}}{4}
\sigma^{(0)}_{\mu}\sigma^{(\2,\alpha)}_{\mu}\right)
\nonumber \\
&+\sum^{N}_{\beta=1}
\left(\sum_{\mu=x,y,z}\frac{J_{13}}{4}\sigma^{(0)}_{\mu}\sigma^{(\3,\beta)}_{\mu}\right),
\label{eq:hamiltonian2}
\end{align}
where $\sigma^{(\2,\alpha)}_{\mu}$ ($\sigma^{(\3,\beta)}_{\mu}$) is 
the $\mu$ component of the Pauli matrices that acts only on the $\alpha$th ($\beta$th) 
qubit in System~II (III).
Let us explain the terms in the above Hamiltonian.
$H_{0}$ makes the quantization axes with the resonance frequencies 
$\omega_{\1}$, $\omega_{\2}$, and $\omega_{\3}$. 
The qubits in System~$\2$ ($\3$) have the same resonance frequency $\omega_{\2}$ ($\omega_{\3}$).
$H^{\rm in}$ represents the interaction between qubits in Systems II and those in III;
there is no interaction among qubits in System~II (III).
$H^{\rm ex}$ is the interactions between System~I and System~II (III).
The qubits in System~II (III) are identically coupled to System~I with the strength $J_{12}$ ($J_{13}$).
We define the initial state similarly to Eq.~(\ref{eq:initial0}):
\begin{align}
	\rho(0)
	&=\frac{1}{2^{L+1}}
	\underbrace{\left(\sigma^{(0)}_{0}+\sigma^{(0)}_{+}+\sigma^{(0)}_{-}\right)}_{\rm System~I}
	\underbrace{\left(\prod^{M}_{\alpha=1}\sigma^{(\2,\alpha)}_{0}\prod^{N}_{\beta=1}
		\sigma^{(\3,\beta)}_{0}\right)}_{{\rm System~II~and~III}}
	\nonumber \\
	&=\frac{1}{2^{L+1}}\left(\sigma^{(0)}_{0}+\sigma^{(0)}_{+}+\sigma^{(0)}_{-}\right),
	\label{eq:initial1}
\end{align}
where $L=M+N$.

The RWA is applied by assuming that the differences 
in the resonance frequencies are significant:
$|\omega_{\1}-\omega_{\2}| \gg J_{12}$, $|\omega_{\2}-\omega_{\3}| \gg J^{(\alpha,\beta)}_{23}$ and 
$|\omega_{\1}-\omega_{\3}| \gg J_{13}$.
We obtain the approximated Hamiltonian in the rotating frame,
\begin{align}
\tilde{H}&\approx \tilde{H}^{\rm in}+\tilde{H}^{\rm ex},
\nonumber\\
\tilde{H}^{\rm in} &= \sum^{M,N}_{\alpha,\beta=1}\frac{J^{(\alpha,\beta)}_{23}}{4}\sigma^{(\2,\alpha)}_{z}\sigma^{(\3,\beta)}_{z},
\nonumber \\
\tilde{H}^{\rm ex}&=\frac{J_{12}}{4}\sigma^{(0)}_{z}\sum^{M}_{\alpha=1}\sigma^{(\2,\alpha)}_{z}
+\frac{J_{13}}{4}\sigma^{(0)}_{z}\sum^{N}_{\beta=1}\sigma^{(\3,\beta)}_{z},
\label{eq:hamiltonian3}
\end{align}
of which the corresponding dynamics from the initial state (\ref{eq:initial1}) is easily solved.
The solution is derived in a quite similar way to Eq.~(\ref{eq:solution0}) as 
\begin{align}
\rho(t) 
&=\frac{\sigma^{(0)}_{0}}{2^{L+1}}
+\frac{\sigma^{(0)}_{+}}{2}\prod^{M}_{\alpha=1}A^{(\2,\alpha)}\prod^{N}_{\beta=1}A^{(\3,\beta)}
\nonumber \\
&+\frac{\sigma^{(0)}_{-}}{2}\prod^{M}_{\alpha=1}A^{(\2,\alpha)}\prod^{N}_{\beta=1}A^{(\3,\beta)},
\nonumber\\
A^{(\2,\alpha)} &=\cos(J_{12}t/2)\frac{\sigma^{(\2,\alpha)}_{0}}{2}
+\sin(J_{12}t/2)\frac{\sigma^{(\2,\alpha)}_{z}}{2}, 
\nonumber \\
A^{(\3,\beta)} &=\cos(J_{13}t/2)\frac{\sigma^{(\3,\beta)}_{0}}{2}
+\sin(J_{13}t/2)\frac{\sigma^{(\3,\beta)}_{z}}{2}.
\label{eq:solution1}
\end{align}
See Appendix \ref{sec:app_a} for the details. 
Note that $J_{23}$ does not influence the dynamics of System~I when the RWA is valid. 
We calculate the following value 
like $S^{\rm TMS}(t)$ in Eq.~\eqref{eq:signalTMS0},
\begin{align}
	S^{\rm TES}(t):=&\tr\left(\sigma^{(0)}_{x}\rho(t)\right)=\left(\cos(J_{12}t/2)\right)^{M}
	\left(\cos(J_{13}t/2)\right)^{N}. 
	\label{eq:signal_TES}
\end{align}

\begin{figure}[b]
	\includegraphics[width=80mm]{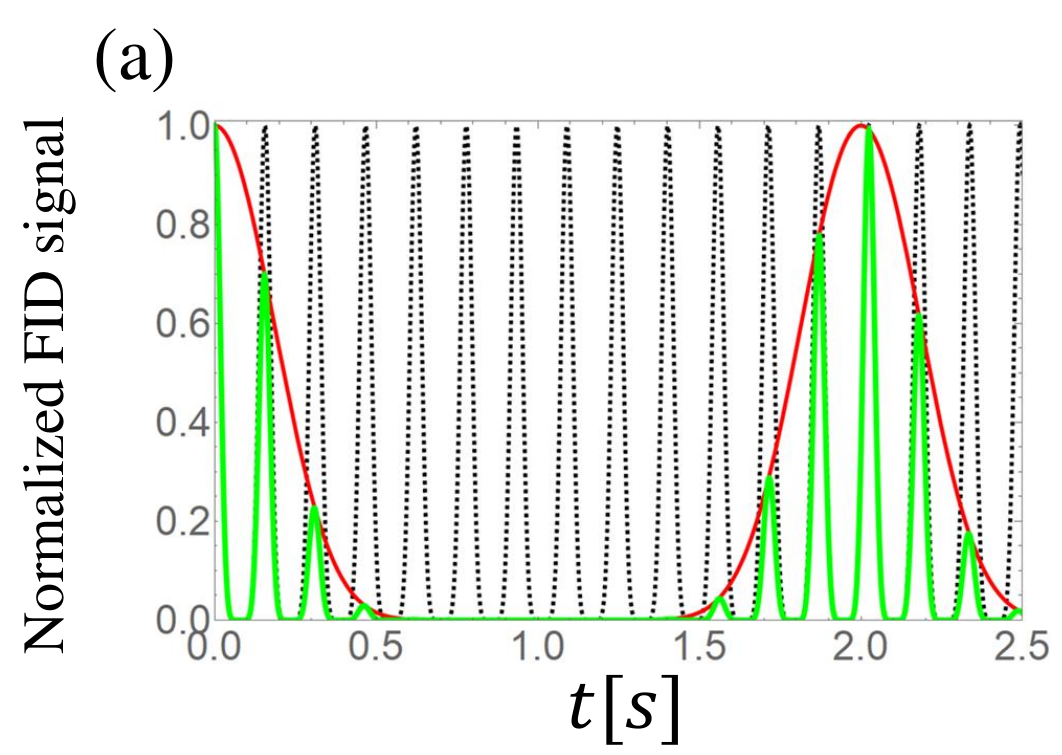}
	
	\includegraphics[width=80mm]{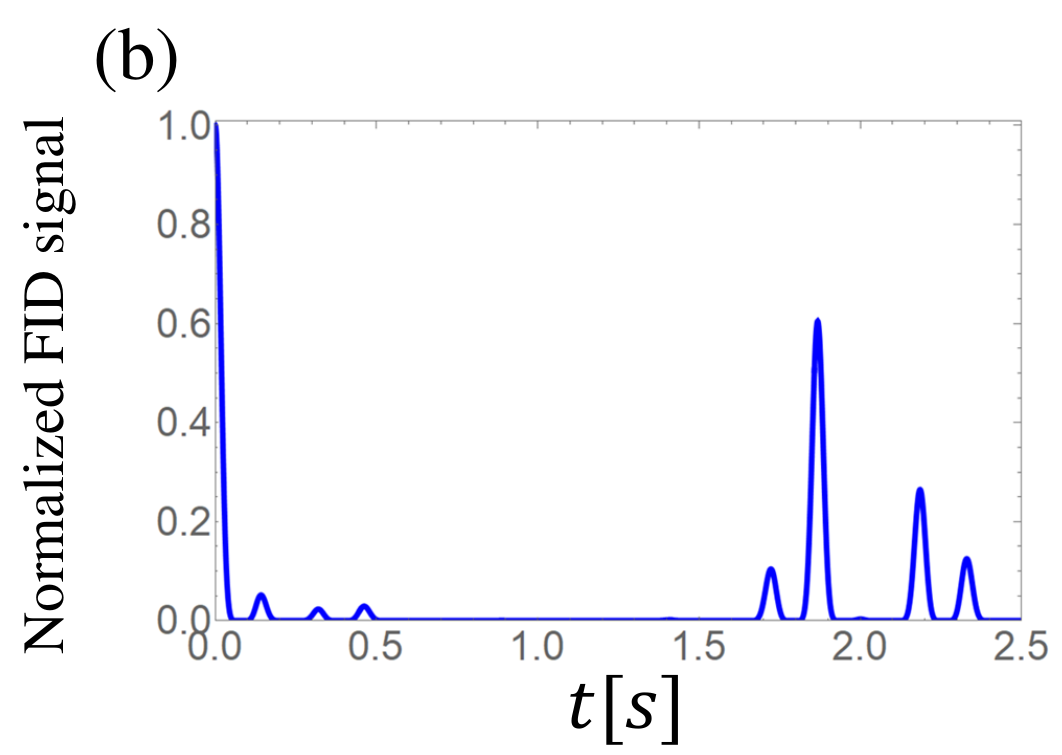}
	\caption{(color online) Suppression of recursions.
	(a) $S^{\rm TES}(t)$ (green line) is plotted.
	For comparison, we also plot $\left( \cos(J_{12} t/2) \right)^{8}$ (dashed black line) and 
	$ \left(\cos( J_{13} t/2  \right)^{12}$	(red line).
	The parameters are taken as $J_{12} = 2\pi \times 6.42$~s$^{-1}$ $J_{13} = 2\pi \times 0.5$~s$^{-1}$ with experimental realization via TES in mind.
	(b) The FID signal $S^{\rm TES}_{\rm virtual}(t)$ of a virtual TES with ${}^{13}$C in the methyl group is plotted.
	The recursion is even more suppressed than $S^{\rm TES}(t)$.}
	\label{fig:TES}
	\end{figure}

We regard Si in the TES molecule as System~I with experimental 
realization in mind. 
Correspondingly, $S^{\rm TES}(t)$ represents the FID signal of Si.
As shown in Fig.~\ref{fig:TMS}(b), we identify the TES molecule with the case of $(M,N)=(8,12)$
by taking H spins in the four methyl groups as System~III.
The qubits in System~II are the nearer H's. 
We again ignore C spins because 
the natural abundance of ${}^{13}$C with a spin half is only 1~\%. 
The parameters are experimentally determined as 
$(J_{12},J_{13},J_{23})
=(2\pi\times 6.42,2\pi\times 0.5,2\pi\times 8.02)$ s$^{-1}$ from 
measured spectra of H and Si. 
We plot $S^{\rm TES}(t)$ (solid green line) with these parameters in Fig.~\ref{fig:TES}~(a).
It shows less recursion than $S^{\rm TMS}(t)$ as expected; 
the peaks of the oscillation with the frequency $J_{12}$ ($J_{13}$) is suppressed due to the other oscillation with $J_{13}$ ($J_{12}$).

To further investigate our idea, we consider a virtual TES molecule by replacing all C atoms  in 
the methyl groups with ${}^{13}$C. 
Although the natural abundance of ${}^{13}$C is only 1~\% and small, we could observe 
the Si spectra peaks caused by ${}^{13}$C's in the normal TES spectrum 
when all H's are decoupled.
We found that the scalar coupling strength $J_{14}$ between Si and ${}^{13}$C in the methyl 
group is $2\pi \times 2.2$~s$^{-1}$ 
(The strength between Si and the other ${}^{13}$C was measured 
as $ 2\pi \times 50.8$~s$^{-1}$).
The corresponding model can be constructed by adding the following term to $\tilde{H}^{\rm ex}$ in Eq. (\ref{eq:hamiltonian3}),
\begin{equation}
\frac{J_{14}}{4}\sigma^{(0)}_{z}\sum^{4}_{\eta=1}\sigma^{({\rm IV},\eta)}_{z},
\end{equation}
where we let System IV represent ${}^{13}$C.
As in the previous case, interactions between C's and H's do not affect the dynamics like $J_{23}$.
We also solve the dynamics in this case and obtain
\begin{equation}
S^{\rm TES}_{\rm virtual}(t)=\left(\cos(J_{12}t/2)\right)^{8}\left(\cos(J_{13}t/2)\right)^{12}\left(\cos(J_{14}t/2)\right)^{4}.
\label{eq:13C}
\end{equation}
As we can see in Fig.~\ref{fig:TES} (b), $S^{\rm TES}_{\rm virtual} (t)$ shows even less recursive behaviour than $S^{\rm TES}(t)$. 
Note that this suppression of the recursion does not occur for $S^{\rm TMS}(t)$ no matter how large $N$ we prepare:
In such a case, each peak will get narrow, but the recursion always occurs at a particular time scale.
This implies that interacting with ancillary DoF with different interaction strengths is essential for realizing a plausible relaxation.
Instead of using labeled ${}^{13}$C, we can employ larger alkyl, such as propyl and butyl, groups.
The corresponding model can be analytically solved even in these cases when the RWA is valid.
	
	The dynamics with dissipation can analytically be solved as well.
	The following terms describe the dissipation:
	\begin{align}
		{\cal L}[\rho]&:={\cal L}^{(0)}+\sum^{M}_{\alpha=1}{\cal L}^{(\2,\alpha)}[\rho]+\sum^{N}_{\beta=1}{\cal L}^{(\3,\beta)}[\rho],
		\nonumber\\
		{\cal L}^{(0)}[\rho]
		&:=\frac{\gamma_{\1}}{2}\left(\sum_{\mu=\pm}\sigma^{(0)}_{\pm}\rho\sigma^{(0)}_{\mp}-\rho\right),
		\nonumber \\
		{\cal L}^{(\2,\alpha)}[\rho]
		&:=\frac{\gamma_{\2}}{2}\left(\sum_{\mu=\pm}\sigma^{(\2,\alpha)}_{\pm}\rho
		\sigma^{(\2,\alpha)}_{\mp}-\rho\right),
		\nonumber\\
		{\cal L}^{(\3,\beta)}[\rho] &:=\frac{\gamma_{\3}}{2}
		\left(\sum_{\mu=\pm}\sigma^{(\3,\beta)}_{\pm}\rho\sigma^{(\3,\beta)}_{\mp}-\rho\right).
		\label{eq:lindbladian1}
	\end{align}
\begin{widetext}
	The solution of the GKSL equation is
	\begin{align}
		\rho(t)&=\frac{1}{2^{L+1}}\sigma^{(0)}_{0}
		+e^{-\gamma_{\1}t/2}\frac{\sigma^{(0)}_{+}}{2}\prod^{M}_{\alpha=1}A_{\cal L}^{(\2,\alpha)}\prod^{N}_{\beta=1}A_{\cal L}^{(\3,\beta)}
		+ e^{-\gamma_{\1}t/2}\frac{\sigma^{(0)}_{-}}{2}\prod^{M}_{\alpha=1}A_{\cal L}^{(\2,\alpha)\dagger}\prod^{N}_{\beta=1}A_{\cal L}^{(\3,\beta)\dagger},
		\nonumber\\
		A_{\cal L}^{(\2,\alpha)}&=e^{-\gamma_{\2} t/2}\bigg(\frac{\gamma_{\2}}{\Omega_{\2}}\sin(\Omega_{\2}t/2)
		+\cos(\Omega_{\2}t/2)\bigg)\frac{\sigma^{(\2,\alpha)}_{0}}{2}
		+e^{-\gamma_{\2} t/2}\bigg( \frac{- i J_{12}}{\Omega_{\2}}\sin(\Omega_{\2}t/2)\bigg)\frac{\sigma^{(\2,\alpha)}_{z}}{2},
		\nonumber\\
		A^{(\3,\beta)} &= e^{-\gamma_{\3} t/2}\bigg(\frac{\gamma_{\3}}{\Omega_{\3}}\sin(\Omega_{\3}t/2)
		+\cos(\Omega_{\3}t/2)\bigg) \frac{\sigma^{(\3,\beta)}_{0}}{2}
		+e^{-\gamma_{\3} t/2}
		\bigg( \frac{- i J_{13}}{\Omega_{\3}}\sin(\Omega_{\3}t/2)\bigg)\frac{\sigma^{(\3,\beta)}_{z}}{2},
		\label{eq:solution1noise}
	\end{align}
	where $\Omega_{\2 (\3)}=\sqrt{J^{2}_{12(13)}-\gamma^{2}_{\2(\3)}}$.
	The expectation value of $\sigma^{(0)}_{x}$ of System~I is given as
	\begin{align}
		S^{\rm TES}_{\cal L}(t) &:=\tr\left(\sigma^{(0)}_{x}\rho_{{\cal L}}(t)\right)
		\nonumber \\
		&=e^{-\gamma_{\1}t/2}\left(e^{-\gamma_{\2} t/2}\bigg(\frac{\gamma_{\2}}{\Omega_{\2}}\sin(\Omega_{\2}t/2)
		+\cos(\Omega_{\2}t/2)\bigg)\right)^{M}
		\left(e^{-\gamma_{\3} t/2}\bigg(\frac{\gamma_{\3}}{\Omega_{\3}}\sin(\Omega_{\3}t/2)
		+\cos(\Omega_{\3}t/2)\bigg)\right)^{N}. 
		\label{eq:signal}
	\end{align}

	We observe the FID signal of Si in a TES molecule and compare it with the function~(\ref{eq:signal}).
	 Figure~\ref{fig:TESwithEXP} shows a qualitative agreement between the model and the corresponding
	  experiment. In the long-time region, the plotted $S^{\rm TES}_{\cal L}(t)$ approximates the experimental result. In contrast, $S^{\rm TES}(t)$ has the revival behavior, which does not appear 
	  in the experiment. Meanwhile, in the short time region, both of $S^{\rm TES}(t)$ and 
	  $S^{\rm TES}_{\cal L}(t)$ well approximate the experimental dynamics; that is, the 
	  relaxation-like behavior of the signal in this region is governed by the interaction with the artificial 
	  environment (H's) rather than the intrinsic one originated from magnetic impurities in solvent.

	\begin{figure}[H]
\centering
	\includegraphics[width=80mm]{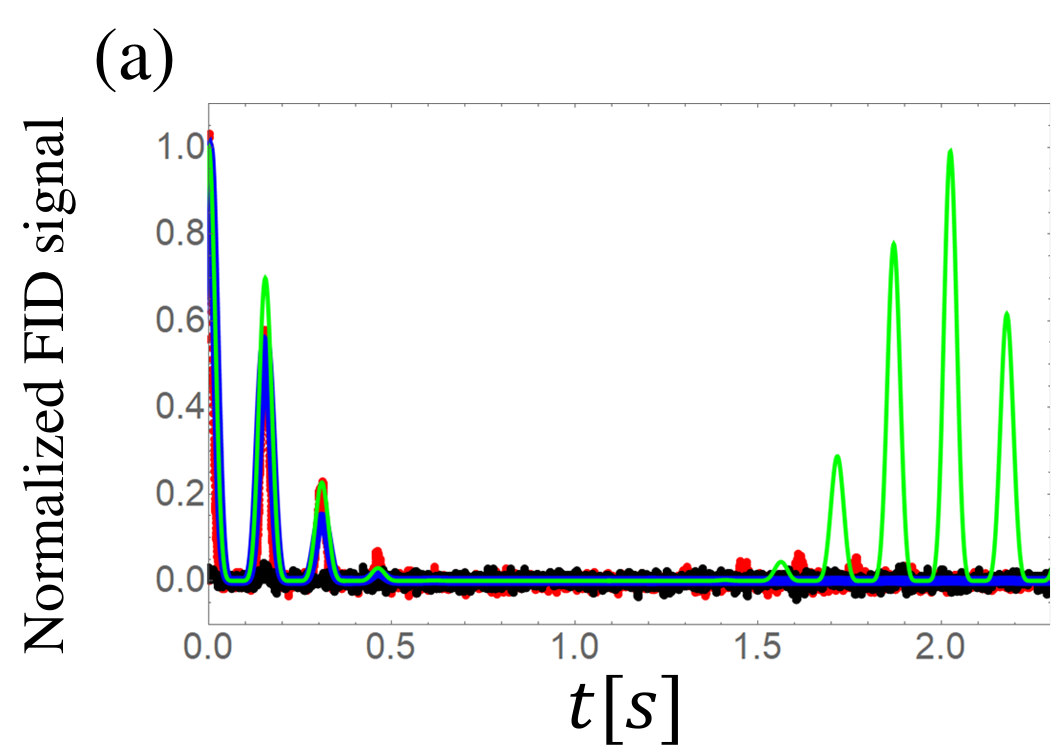}
	\hspace{1cm}
	\includegraphics[width=80mm]{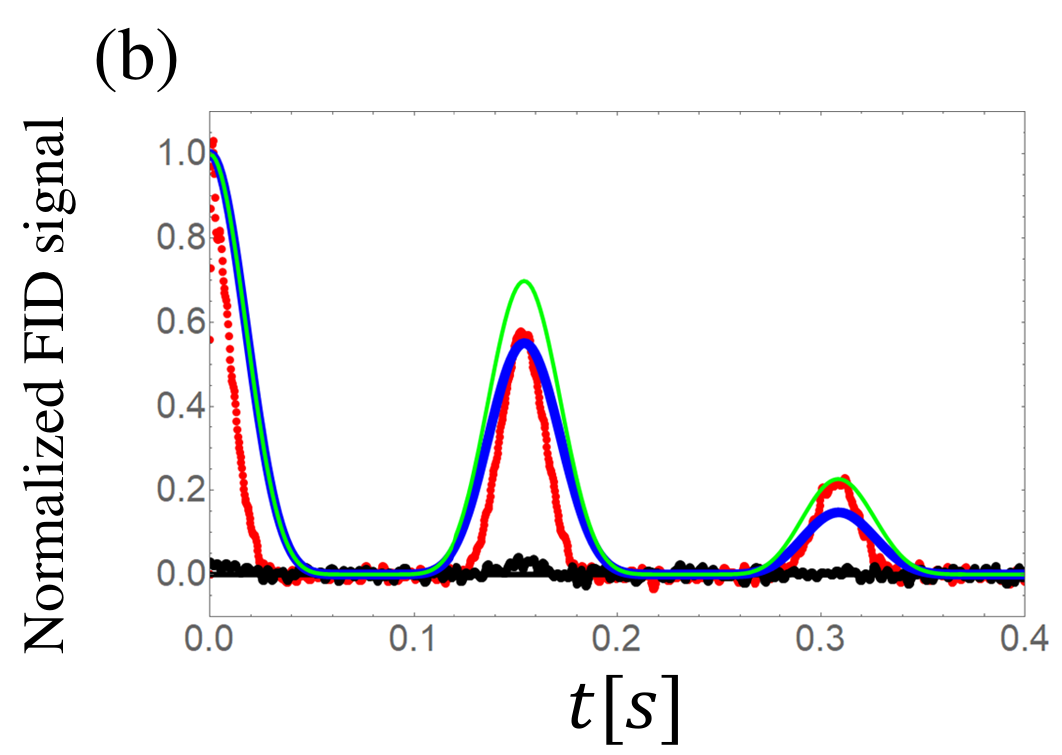}
	\caption{(color online) The dynamics of $S^{\rm TES}(t)$, $S^{\rm TES}_{\cal L}(t)$, and the corresponding
	experimental result in (a) a long time region $t \in[0,2.5]$~s, and (b) a short time region 
	$t \in [0,0.4]$~s. In both panels, the green line represents 
	$S^{\rm TES}(t)$ 
	while the blue line $S^{\rm TES}_{\cal L}(t)$. The blue line in (a) is not visible because all lines 
	overlap with each other. 
	The red dots are the experimental result of the (real part of) FID signal of Si.
	The black dots are the imaginary part of the FID signal, which should be taken 
	$0$ in this situation.
	We need to compare $S^{\rm TES}(t)$, $S^{\rm TES}_{\cal L}(t)$ and the red dots.}
	\label{fig:TESwithEXP}
	\end{figure}

\end{widetext}

\subsection{Case when Rotating Wave Approximation is not applicable}
\label{NRWA}

Connecting larger alkyl groups to Si than TES 
will make a more plausible artificial environment.
Its dynamics can be solved easily with the method discussed in Appendix~A.
However, it is not easy to prepare the corresponding molecule.
they are not usually available. 
In this subsection, instead, we introduce another way to increase the environmental DoF;
we implement complicated interactions by quitting to apply the RWA 
(weak coupling limit $\rightarrow$ strong coupling region \cite{Levitt2008}), or equivalently, 
by decreasing an applied magnetic field in the case of NMR experiments.

\begin{figure}[t]
	\begin{center}
		\includegraphics[width=80mm]{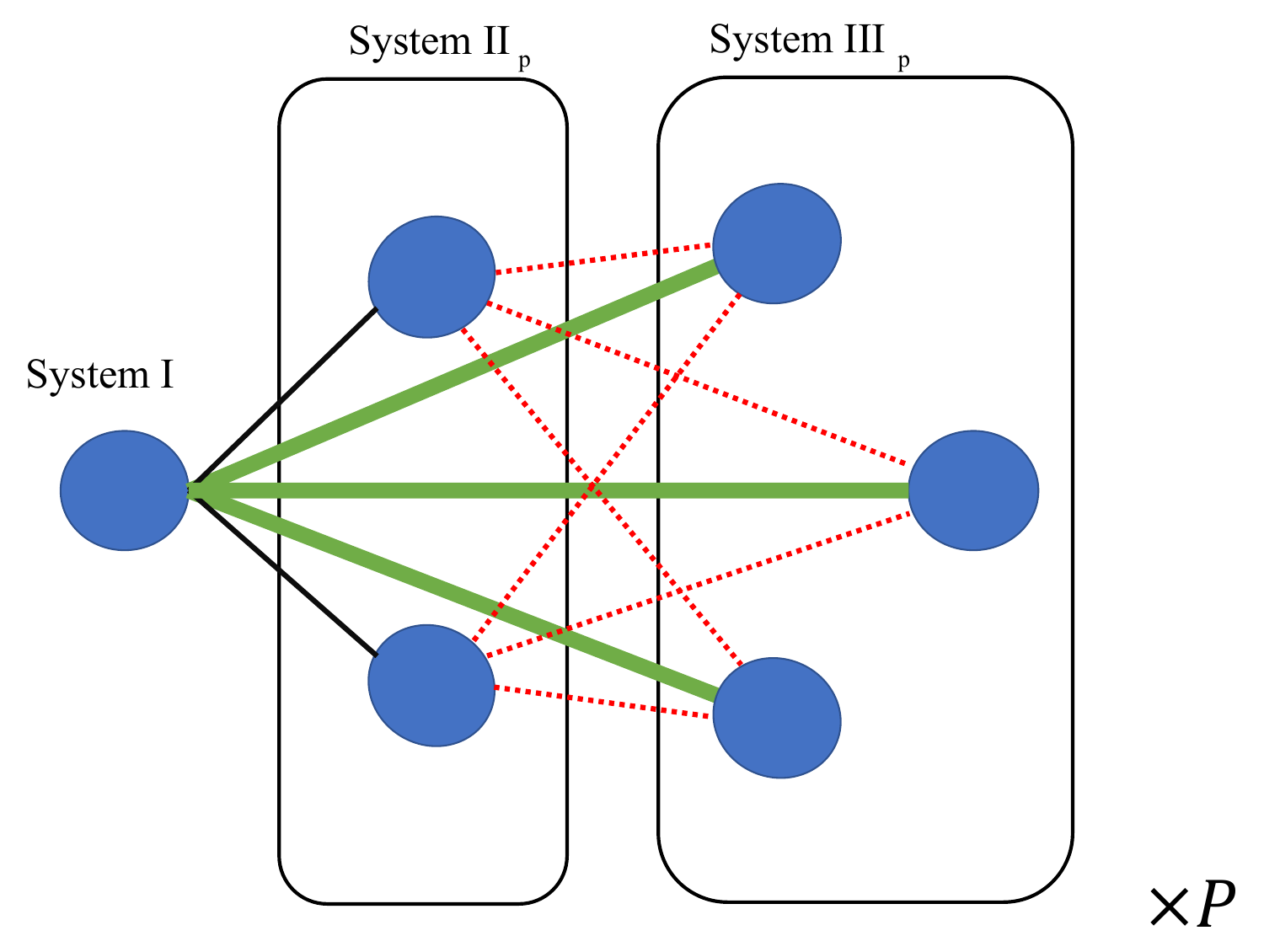}
		\caption{(color online) 
			System~I interacts with  System~II and III, divided into identical $P$ parts.
			One of them is shown here. 
			System~I interacts with System~II$_{p}$ (black thin lines) 
			and with System~III$_{p}$ (green thick lines).
			Also, qubits in System~II and III interact with each other (red thin dotted lines). 
			There is neither interaction among qubits in System~II$_{p}$ (III$_{p}$) nor inter-parts ones. 
			\label{fig:opensystem}
		}
	\end{center}
\end{figure}

We reconsider the Hamiltonian (\ref{eq:hamiltonian2}).
For later convenience, we decompose both of System~II and III into identical $P$ parts: System~II$_{p}$ and III$_{p}$ ($p=1,\cdots,P$), see Fig.~\ref{fig:opensystem}.
Each System~II$_{p}$ (III$_{p}$) is comprised of $m$ ($n$) qubits;
accordingly, System~II (III) has $M=Pm$ ($N=Pn$) qubits in total.
The qubits in System~II$_{p}$ and III$_{p}$ have an interaction 
with the identical interaction strength $J_{23}$ while there is no interaction 
among qubits in System~II$_{p}$ (III$_{p}$).
Also, we assume that System ~II$_{p}$ (III$_{p}$) does not interact with System ~II$_{p'}$ (III$_{p'}$) for $p'\neq p$.
The Hamiltonian is rearranged as
\begin{align}
	H &=H_{0}+\sum^{P}_{p=1}\left(H^{\rm in}_{p}+H^{\rm ex} _{p}\right),
	\nonumber \\
	H_{0}  &=\frac{\omega_{\1}}{2}\sigma_{z}^{(0)}
	+\frac{\omega_{\2}}{2}\sum_{p=1}^{P}\sum^{m}_{\alpha=1}\sigma^{(\2_{p},\alpha)}_{z}
	\nonumber \\
	&+\frac{\omega_{\3}}{2}\sum_{p=1}^{P}\sum^{n}_{\beta=1}\sigma^{(\3_{p},\beta)}_{z},
	\nonumber\\
	H^{\rm in}_{p} &=\sum^{m,n}_{\alpha,\beta=1}\sum_{\mu=x,y,z}
	\frac{J_{23}}{4}\sigma^{(\2_{p},\alpha)}_{\mu}\sigma^{(\3_{p},\beta)}_{\mu},
	\nonumber \\
	H^{\rm ex}_{p} &=\sum^{m}_{\alpha=1}\left(\sum_{\mu=x,y,z}\frac{J_{12}}{4}
	\sigma^{(0)}_{\mu}\sigma^{(\2_{p},\alpha)}_{\mu}\right)
	\nonumber \\
	&+\sum^{n}_{\beta=1}
	\left(\sum_{\mu=x,y,z}\frac{J_{13}}{4}\sigma^{(0)}_{\mu}\sigma^{(\3_{p},\beta)}_{\mu}\right),
	\label{eq:hamiltonian5}
\end{align}
where $\sigma^{(\2_{p},\alpha)}_{\mu}$ ($\sigma^{(\3_{p},\beta)}_{\mu}$) is 
the $\mu$ component of the Pauli matrices that acts only on the $\alpha$th ($\beta$th) 
qubit in System~II$_{p}$ (III$_{p}$).

We also rearrange the initial state~(\ref{eq:initial1}) as
\begin{align}
	\rho(0)
	&=\frac{1}{2^{L+1}}
	\underbrace{\left(\sigma^{(0)}_{0}+\sigma^{(0)}_{+}+\sigma^{(0)}_{-}\right)}_{\rm System~I}
	\nonumber \\
	&\times \prod^{P}_{p=1}\underbrace{\left(\prod^{m}_{\alpha=1}\sigma^{(\2_{p},\alpha)}_{0}\prod^{n}_{\beta=1}
		\sigma^{(\3_{p},\beta)}_{0}\right)}_{p{\rm th~part~of~System~II~and~III}}
	\nonumber \\
	&=\frac{1}{2^{L+1}}\left(\sigma^{(0)}_{0}+\sigma^{(0)}_{+}+\sigma^{(0)}_{-}\right),
	\label{eq:initial2}
\end{align}
where $L$ denotes $P(m+n)$.

\begin{figure}[h]
	\begin{center}
		\includegraphics[width=80mm]{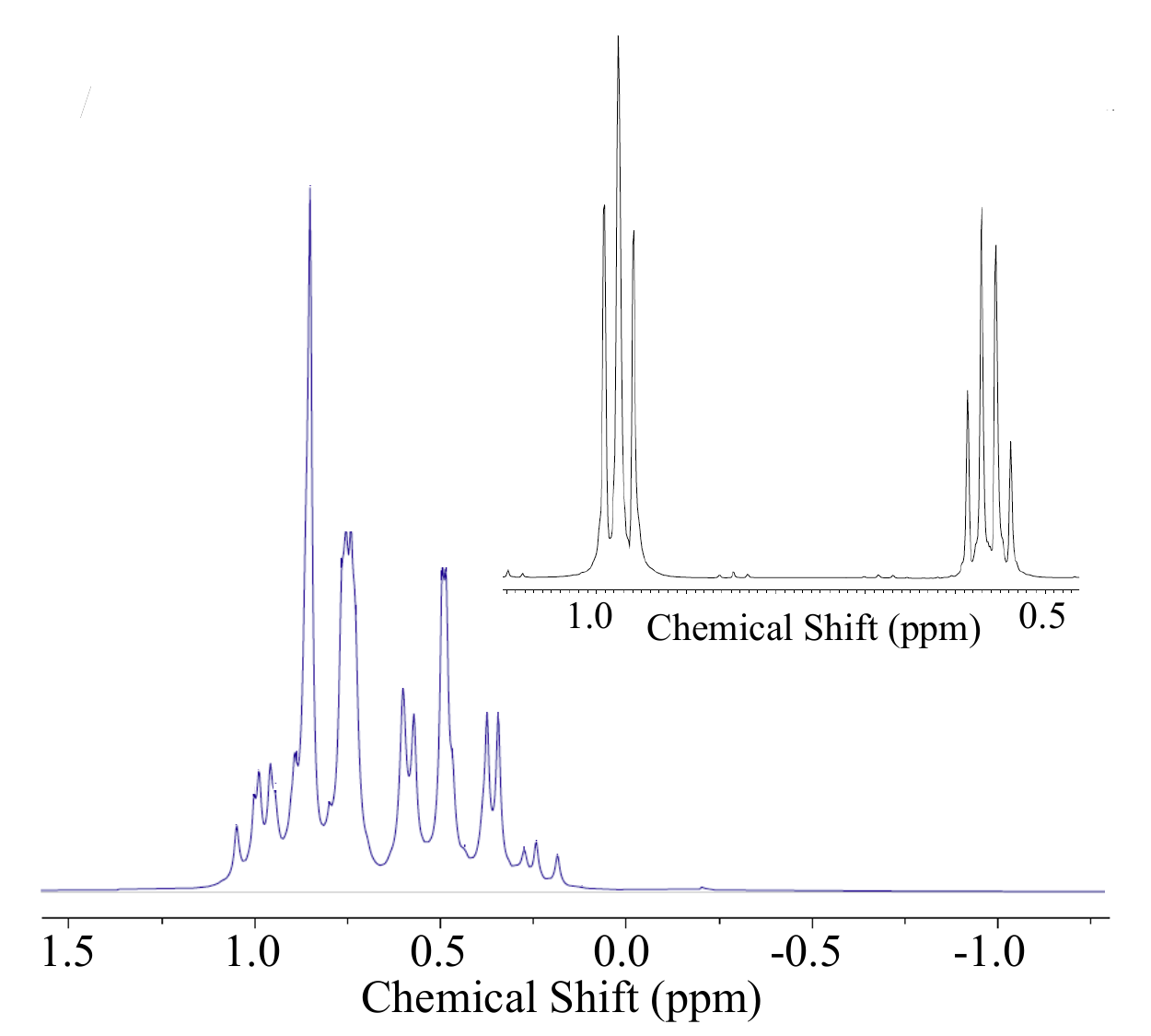}
		\caption{The spectrum of H of TES at 1.4~T. The inset spectrum is 
			taken at 11.7~T. The peaks corresponding to H's in System~II and 
			those in System~III are well separated 
			at 11.7~T, but they are not at 1.4~T. 
			\label{fig:HLowTES}
		}
	\end{center}
\end{figure}

In the case of the TES molecule, the parameters are taken as $(P,m,n)=(4,2,3)$.Each $p$ part corresponds to an alkyl group out of four (a part has no interaction with the other three parts).If we take the RWA for all interactions, the Hamiltonian (\ref{eq:hamiltonian3}) is reproduced.Note that $| \omega_{\2} -\omega_{\3}| /J_{23} \sim 26$ in the experiments at 11.7~T of the magnetic field strength, shown in \S~\ref{RWA}, and thus we can employ the RWA for them (Fig.~\ref{fig:HLowTES}).

In NMR, the resonance frequencies $\omega_{\1}$, $\omega_{\2}$, and $\omega_{\3}$ 
are proportional to a common magnetic field $B$, where the coefficients are different 
gyromagnetic ratios.
Remembering the case of the TES molecule, the difference between the gyromagnetic 
ratios in $\omega_{\2}$ and $\omega_{\3}$ will not be so large because the qubits 
in System~II and III are homonuclear (H).
Thus, the RWA for $H^{\rm in}_{p}$ will be valid only when $B$ is sufficiently large
as in the case of the experiments in \S~\ref{RWA}.  
On the other hand, the RWA for $H^{\rm ex}_{p}$ does not require a large $B$ 
because it is an interaction between heteronuclear Si and H's.
Summarizing, we can consider the situation where 
$\omega_{\1},\omega_{\2},\omega_{\3} \gg 0$, $|\omega_{\1}-\omega_{\2}| \gg J_{12}$ and
$|\omega_{\1}-\omega_{\3}| \gg J_{13}$, but 
$\delta \omega:= |\omega_{\2}-\omega_{\3} | \sim J_{23}$.
By adjusting the strength of $B$, we can implement the following Hamiltonian 
in the case of the TES molecule:
\begin{widetext}
\begin{align}
	\tilde{H}' &\approx\sum^{P}_{p=1}\left(\tilde{H}^{\rm in}_{p}+\tilde{H}^{\rm ex}_{p}\right),
	\nonumber\\
	\tilde{H}^{\rm in}_{p}(t) &=
	\sum^{m,n}_{\alpha,\beta=1}\left(\frac{J_{23}}{4}\sigma^{(\2_{p},\alpha)}_{z}\sigma^{(\3_{p},\beta)}_{z}
	+\frac{J_{23}}{2}\left(e^{i \delta \omega t}\sigma^{(\2_{p},\alpha)}_{+}\sigma^{(\3_{p},\beta)}_{-}
	+e^{-i \delta \omega t}\sigma^{(\2_{p},\alpha)}_{-}\sigma^{(\3_{p},\beta)}_{+}\right)\right),
	\nonumber\\
	\tilde{H}^{\rm ex}_{p}=&\sum^{m}_{\alpha=1}\frac{J_{12}}{4}\sigma^{(0)}_{z}\sigma^{(\2_{p},\alpha)}_{z}
	+\sum^{n}_{\beta=1}\frac{J_{13}}{4}\sigma^{(0)}_{z}\sigma^{(\3_{p},\beta)}_{z}.
\end{align}
Consider noiseless dynamics from the initial state (\ref{eq:initial2}).
Let us take an ansatz that the solution has the following form:
\begin{align}
	\rho(t)=&\frac{\sigma^{(0)}_{0}}{2^{L+1}}
	+\frac{\sigma^{(0)}_{+}}{2}\prod^{P}_{p=1}C_{p}(t)
	+\frac{\sigma^{(0)}_{-}}{2}\prod^{P}_{p=1}C^{\dagger}_{p}(t).
\end{align}
We easily justify the above ansatz and obtain the dynamical equation for $C_{p}(t)$ as
\begin{equation}
	\frac{d C_{p}(t)}{d t}=-i\Bigl\{ \frac{J_{12}}{4}\sum^{m}_{\alpha=1}\sigma^{\rm (II_{p},\alpha)}_{z}+\frac{J_{13}}{4}\sum^{n}_{\beta=1}\sigma^{\rm (III_{p},\beta)}_{z}, C_{p}(t)\Bigr\}-i \Bigl[\tilde{H}^{\rm in}_{p}(t) , C_{p}(t)\Bigr].
	\label{eq:dynamics3}
\end{equation}
The operation $\{\bullet,\bullet\}$ represents an anti-commutator, which comes from the relation $\sigma_{z}\sigma_{+}=-\sigma_{+}\sigma_{z}=\sigma_{+}$.
Similarly, dynamics with the intrinsic environment can be considered with the following ansatz:
\begin{align}
	\rho_{\cal L}(t)=&\frac{\sigma^{(0)}_{0}}{2^{L+1}}
	+e^{-\gamma_{\1}t/2}\frac{\sigma^{(0)}_{+}}{2}\prod^{P}_{p=1}C^{\cal L}_{p}(t)
	+e^{-\gamma_{\1}t/2}\frac{\sigma^{(0)}_{-}}{2}\prod^{P}_{p=1}C^{{\cal L}\dagger}_{p}(t).
\end{align}
The corresponding equation is
\begin{align}
	\frac{d C^{\cal L}_{p}(t)}{d t}=&-i\Bigl\{ \frac{J_{12}}{4}\sum^{m}_{\alpha=1}\sigma^{\rm (II_{p},\alpha)}_{z}+\frac{J_{13}}{4}\sum^{n}_{\beta=1}\sigma^{\rm (III_{p},\beta)}_{z}, C^{\cal L}_{p}(t)\Bigr\}-i \Bigl[\tilde{H}^{\rm in}_{p} (t), C^{\cal L}_{p}(t)\Bigr]
	+\sum^{m}_{\alpha=1}{\cal L}^{(\2_{p},\alpha)}[C^{\cal L}_{p}(t)]+\sum^{n}_{\beta=1}{\cal L}^{(\3_{p},\beta)}[C^{\cal L}_{p}(t)].
	\label{eq:dynamics4}
\end{align}
\end{widetext}
According to the above equations~(\ref{eq:dynamics3}) and (\ref{eq:dynamics4}), $C_{p}(t)$ ($C^{\cal L}_{p}(t)$) are identical for any $p=1,\cdots,P$, and thus we let $C(t)$ ($C^{\cal L}(t)$) be the solution of 
the above equations.
Unlike in \S~\ref{RWA}, $\tilde{H}^{\rm in}_{p}$ does not allow us to represent the solution in a simple form.
The expectation value of $\sigma^{(0)}_{x}$ is given as
\begin{align}
	S^{\rm TES'}(t) &:=\tr\left(\sigma^{(0)}_{x}\rho(t)\right)=\left(\tr \left({\rm Re}~C(t)\right)\right)^{P},
	\nonumber \\
	S^{\rm TES'}_{\cal L}(t ) &:=\tr\left(\sigma^{(0)}_{x}\rho_{\cal L}(t)\right)=\left(\tr\left({\rm Re}~C^{\cal L}(t)\right)\right)^{P}.
	\label{eq:signal2}
\end{align}
The solution $C(t)$ has more DoF than the model in the RWA, although 
it is difficult to solve the dynamical equation analytically.
This large freedom leads to faster relaxation-like behavior and less recursive behavior. 

\begin{figure}[H]
\begin{tabular}{cc}
\begin{minipage}[t]{0.9\hsize}
\centering
\includegraphics[width=80mm]{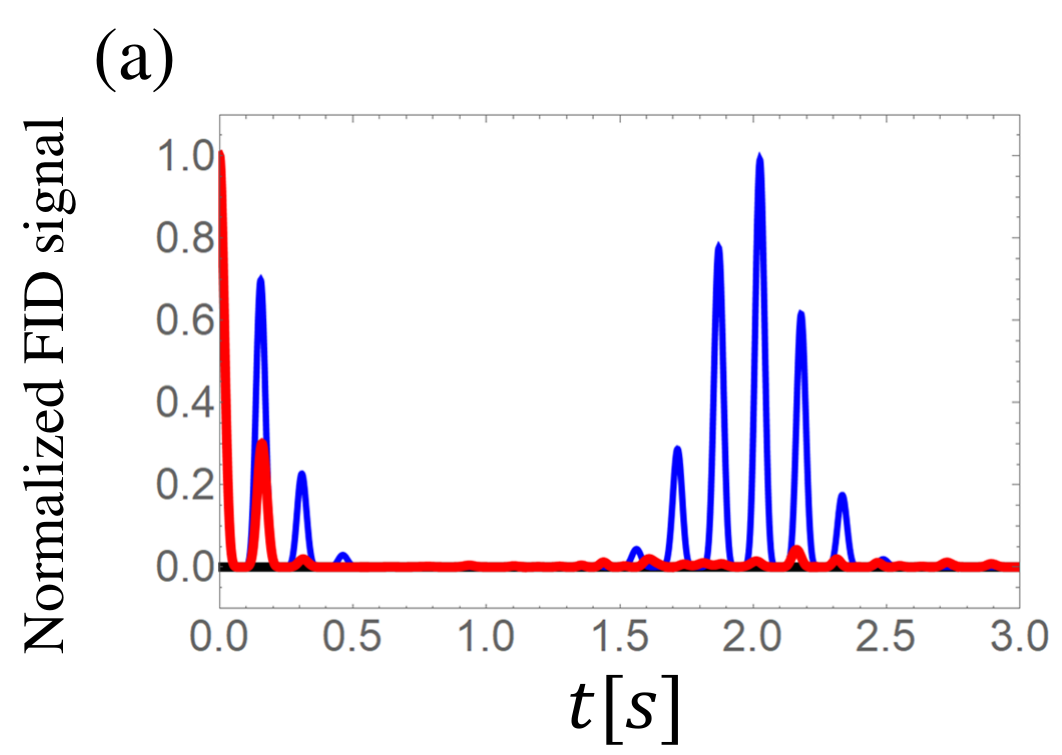}
\end{minipage} \\
\begin{minipage}[t]{0.9\hsize}
\centering
\includegraphics[width=80mm]{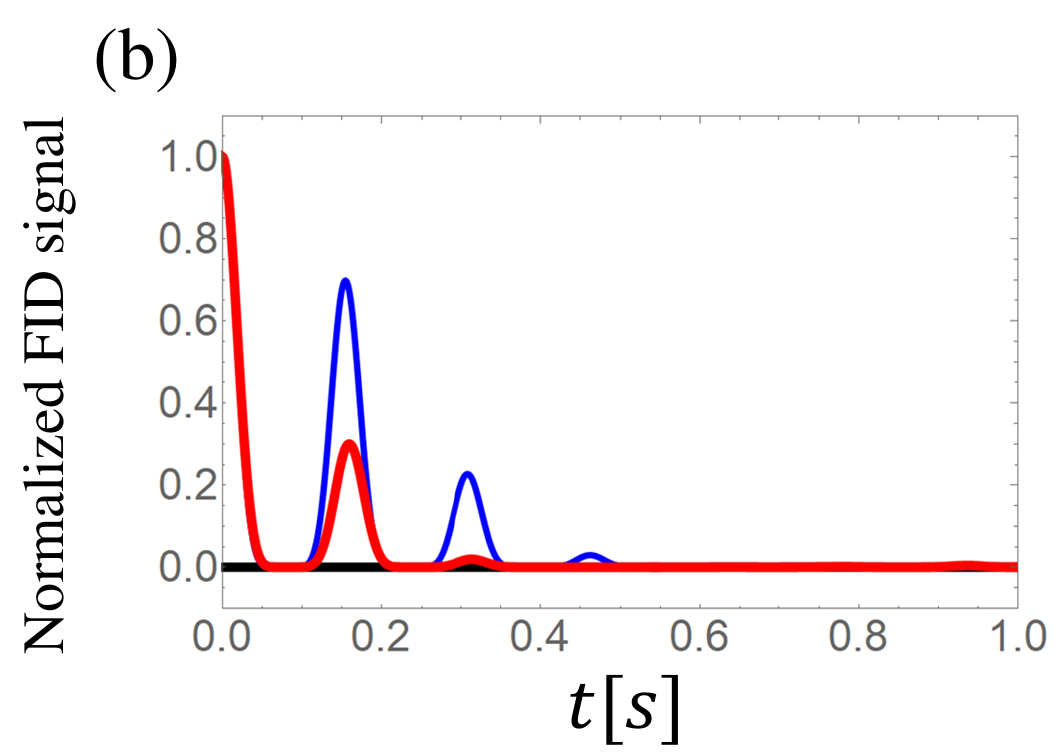}
\end{minipage}
\end{tabular}
\caption{(color online) The comparison of $S^{\rm TES}(t)$ and $S^{\rm TES'}(t)$ for the noiseless case in (a) the long time region $t \in [0, 3.0]$~s and 
(b) short time region $t \in [0, 1.0]$~s. The red line represents 
$S^{\rm TES'}(t)$, the blue one $S^{\rm TES}(t)$.}
\label{fig:xxwithzz}
\end{figure}

We first compare $S^{\rm TES}(t)$ and $S^{\rm TES'}(t)$ (Fig.~\ref{fig:xxwithzz}).
The parameter $\delta \omega$ is taken to be $2\pi \times 24.8$~s$^{-1}$
in order to compare the results with the corresponding experiment later.
In Fig.~\ref{fig:xxwithzz}(a), we see that the recursive behaviour 
in $S^{\rm TES}(t)$ is suppressed in $S^{\rm TES'}(t)$.
The additional interaction in $\tilde{H}^{\rm in}_{p}$ increases effective DoF during the dynamics 
and results in this suppression.
As shown in Fig.~\ref{fig:xxwithzz}(b), the relaxation-like behaviour in $S^{\rm TES'}(t)$ 
is quite faster than that in $S^{\rm TES}(t)$.
Even the third peak can hardly be observed in $S^{\rm TES'}(t)$.
Note that these behaviors originate from the interaction with the H nuclei because the dissipation term 
is not considered now.

\begin{figure}[htb]
\begin{tabular}{cc}
\begin{minipage}[t]{0.9\hsize}
\centering
\includegraphics[width=80mm]{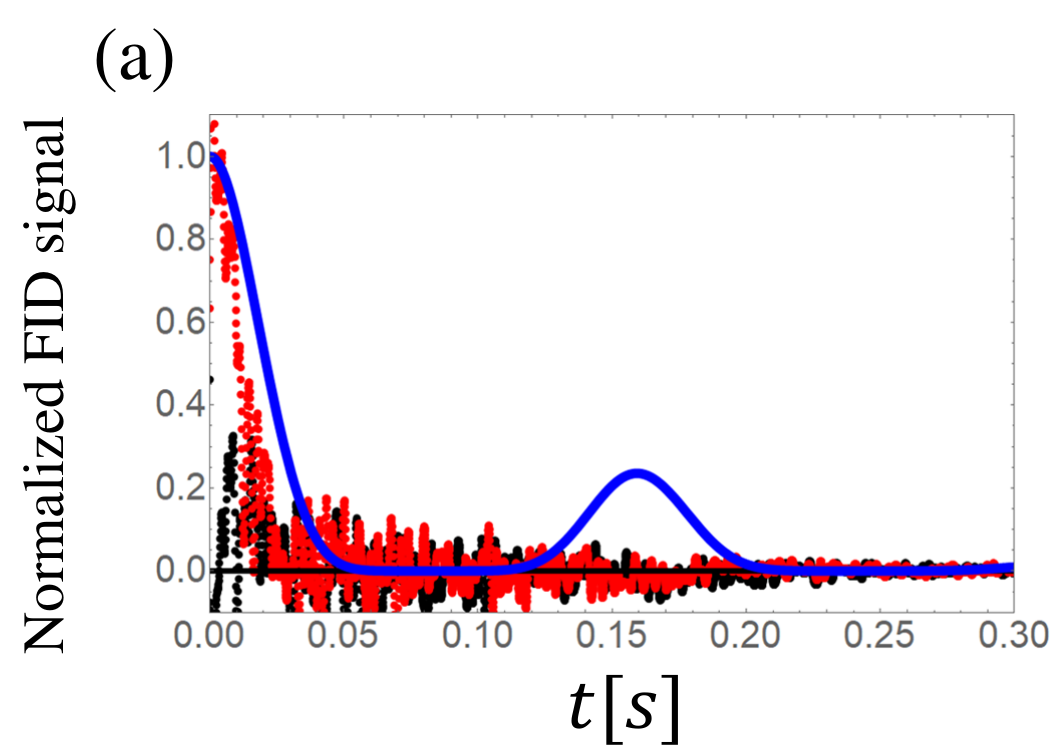}
\end{minipage} \\
\begin{minipage}[t]{0.9\hsize}
\centering
\includegraphics[width=80mm]{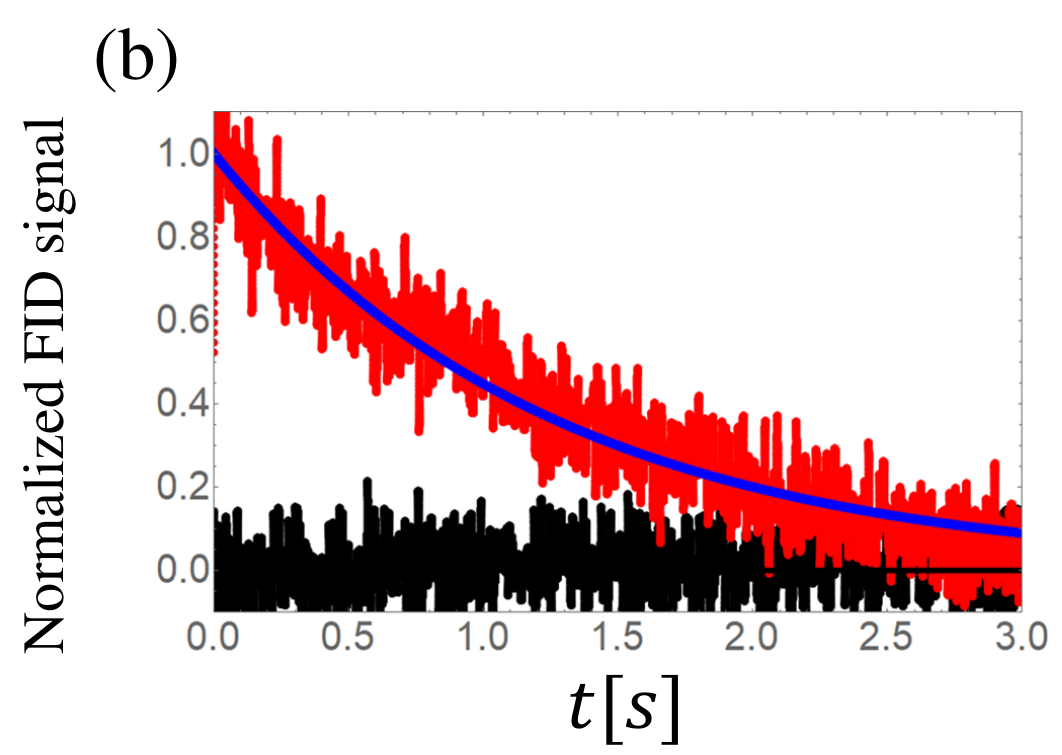}
\end{minipage}
\end{tabular}
\caption{(color online) Measured FID signal (a) without and (b) with decoupling all H's in TES molecules. 
The red dots are the experimental result of the (real part of) the FID signal of the Si. 
The black dots are the imaginary part of the FID signal, which should be taken $0$ 
in this situation.
From (b), we find that $\gamma_{\1} \sim 2^{-1} $~s$^{-1}$, which cannot explain the faster experimental relaxation in (a).}
\label{fig:xxwithEXP}
\end{figure}

We observed the FID signal of Si in TES 
under a small magnetic field $|B|=1.4$~T (60~MHz of H resonance frequency, Fig.~\ref{fig:HLowTES}), 
which gives $\delta \omega :=|\omega_{\2}-\omega_{\3}|= 2\pi \times 24.8$~s$^{-1}$,
as shown in Fig.~\ref{fig:xxwithEXP}~(a) with the theoretically 
predicted signal $S^{\rm TES'}_{\cal L}(t)$.
The FID signal is consistent with our calculation in $t \in [0, 0.1]$ although it is noisy:
The sensitivity of a FID signal is known to be proportional to $B^{2 \sim 3}$\cite{Levitt2008,Claridge}. 
On the other hand, we observe no recursive behavior theoretically predicted in $t \in [0.1, 0.2]$.  
We confirmed that this behavior is not caused by $\gamma_{\1}$,
which represents direct relaxation due to the intrinsic environment, 
by observing the FID signal when the DoF of $H$ is decoupled. 
We used a standard NMR technique \cite{Levitt2008,Claridge} to perform this procedure.
In Fig.~\ref{fig:xxwithEXP}(b), we observed that $\gamma_{\1}$ 
is not large, therefore the no-recursive behavior in Fig.~\ref{fig:xxwithEXP}(a) 
cannot be explained by the effect of $\gamma_\1$. 

It seems that the model may be over-simplified:
The model includes two assumptions, (1) there is no interaction among qubits in System~II$_{p}$ (III$_{p}$)
and (2)  System ~II$_{p}$ (III$_{p}$) does not interact with System ~II$_{p'}$ (III$_{p'}$) for $p'\neq p$. Therefore, the FID signal exhibited more plausible ``relaxation-like'' behavior than 
we calculated because of unexpected more DoF.

\section{summary}
\label{sec:conclusion}

We have extended the open system model \cite{Ho_2019} to apprehend the concept of open systems. 
This model comprises System~I (a single qubit), System~II (qubits interacting 
with System~I), and an intrinsic environment that weakly interacts with System~I and II. 
We focus on the phase dynamics of System~I, which is described with the GKSL equation. 
This dynamics is influenced by the intrinsic environment in a long time scale, but is approximately isolated from it in a short time scale: The dynamics in the short time
scale is almost determined by the only interaction between System~I and II,
and exhibits a relaxation-like behavior. Therefore, System~II acts as an artificial 
environment for System~I in the short time scale. 
The dynamics exhibits a recursive behavior because System~II's degrees of freedom 
(DoF) are few. However, we observe ``the more DoF, the less recursive behavior''. 
This model is easily implemented 
with a proper molecule, such as Tetramethylsilane (TMS) and Tetraethylsilane (TES),   
using standard high precision NMR equipment for chemical analysis. 
The experimental results have a good agreement with the theoretical ones.

\section*{Acknowledgment}

We are grateful to  Hiroyuki Sugihara in JASCO INTERNATIONAL CO., LTD., 
who performed the experiment shown in \S~\ref{NRWA}.
This work was supported by JSPS Grants-in-Aid for Scientific Research 
(21K03423) and CREST (JPMJCR1774).

\appendix
\section{Derivation of Eqs.~(\ref{eq:solution0}), (\ref{eq:solution0noise}), (\ref{eq:solution1}), and (\ref{eq:solution1noise})}
\label{sec:app_a}

To derive the analytical solutions~(\ref{eq:solution0}), (\ref{eq:solution0noise}), (\ref{eq:solution1}), and (\ref{eq:solution1noise}),
it is enough to consider the dynamics with the following Hamiltonian and dissipation terms:
\begin{align}
H&=\sum^{N}_{k=1}\frac{J_{k}}{4}\sigma^{(0)}_{z}\sigma^{(k)}_{z}
+\sum^{N}_{k,k'=1}\frac{J_{k,k'}}{4}\sigma^{(k)}_{z}\sigma^{(k')}_{z}
\nonumber \\
&=\sum^{N}_{k=1}H^{\rm ex}_{k}+\sum^{N}_{k,k'=1}H^{\rm in}_{k,k'},\nonumber\\
{\cal L}[\rho]=&\sum^{N}_{i=0}{\cal L}^{(i)}[\rho]=\sum^{N}_{i=0}\frac{\gamma_{i}}{2}\left(\sum_{\mu=\pm}
\sigma^{(i)}_{\pm}\rho\sigma^{(i)}_{\mp}-\rho\right)
\label{eq:app_dynamics}
\end{align}
with the initial state~(\ref{eq:initial0}).
The parameters $J_{k}$ ($k=1,\cdots,N$) and $\gamma_{i}$ ($i=0,1,\cdots,N$) can be different depending on the qubit.
The latter term in the Hamiltonian represents an interaction between System~II and III in the main text or, in general, interactions among them.
By taking appropriate parameters $(J_{k},\gamma_{i},J_{k,k'})$, we reproduce the Hamiltonian and dissipation terms (\ref{eq:hamiltonian0}), (\ref{eq:lindbladian0}), (\ref{eq:hamiltonian3}), and (\ref{eq:lindbladian1}).

\begin{widetext}
We set an ansatz that the state during the dynamics has the form,
\begin{align}
\rho_{\cal L}(t)=&\frac{\sigma^{(0)}_{0}}{2^{N+1}}
+e^{-\gamma_{0}t/2}\frac{\sigma^{(0)}_{+}}{2}\prod^{N}_{k=1}
D_{\cal L}^{(k)}(t)+e^{-\gamma_{0}t/2}\frac{\sigma^{(0)}_{-}}{2}\prod^{N}_{k=1}
D_{\cal L}^{(k)\dagger}(t),\nonumber\\
D_{\cal L}^{(k)}(t)=&d^{(k)}_{0}(t)\frac{\sigma^{(k)}_{0}}{2}+d^{(k)}_{z}(t)
\frac{\sigma^{(k)}_{z}}{2}.
\end{align}
The initial state~(\ref{eq:initial0}) corresponds to $\left(d^{(k)}_{0}(0),d^{(k)}_{z}(t)\right)=(1,0)$.
Assuming this initial state, we can easily show that the above ansatz is justified.
Therefore, the dynamical variables are now $\{d^{(k)}_{0}(t),d^{(k)}_{z}(0)\}^{N}_{k=1}$.
As the ansatz is justified, the latter part $H^{\rm in}_{k,k'}$ in the Hamiltonian does not affect the dynamics:
$\rho_{\cal L}(t)$ always commutes with this interaction Hamiltonian during the dynamics.
The GKSL equation is evaluated as
\begin{align}
{\cal G}[\rho_{\cal L}(t)] &:=-i[H,\rho_{\cal L}(t)]+{\cal L}[\rho_{\cal L}(t)]
\nonumber\\
&=e^{-\gamma_{0}t/2}\sum^{N}_{k=1}D_{\cal L}^{(1)}(t)\cdots\left(-i\left[H^{\rm ex}_{k}, \frac{\sigma^{(0)}_{+}}{2}D_{\cal L}^{(k)}(t)\right]\right)\cdots D_{\cal L}^{(N)}(t)
\nonumber\\
&+e^{-\gamma_{0}t/2}\sum^{N}_{k=1}D_{\cal L}^{(1)\dagger}(t)\cdots 
\left(-i\left[H^{\rm ex}_{k},\frac{\sigma^{(0)}_{-}}{2} D_{\cal L}^{(k)\dagger}(t)\right]\right)\cdots D_{\cal L}^{(N)\dagger}(t)
\nonumber\\
&+e^{-\gamma_{0}t/2}{\cal L}^{(0)}\left[\frac{\sigma^{(0)}_{+}}{2}\right]\prod^{N}_{k=1}
D_{\cal L}^{(k)}(t)+e^{-\gamma_{0}t/2}{\cal L}^{(0)}\left[\frac{\sigma^{(0)}_{-}}{2}\right]\prod^{N}_{k=1}
D_{\cal L}^{(k)\dagger}(t)
\nonumber\\
&+e^{-\gamma_{0}t/2}\frac{\sigma^{(0)}_{+}}{2}\sum^{N}_{k=1}D_{\cal L}^{(1)}(t)\cdots {\cal L}^{(k)}[D_{\cal L}^{(k)}(t)]\cdots D_{\cal L}^{(N)}(t)
\nonumber\\
&+e^{-\gamma_{0}t/2}\frac{\sigma^{(0)}_{-}}{2}\sum^{N}_{k=1}D_{\cal L}^{(1)\dagger}(t)\cdots {\cal L}^{(k)}[D_{\cal L}^{(k)\dagger}(t)]\cdots D_{\cal L}^{(N)\dagger}(t)
\nonumber\\
&=-\frac{\gamma_{0}}{2}\left(e^{-\gamma_{0}t/2}\frac{\sigma^{(0)}_{+}}{2}\prod^{N}_{k=1}
D_{\cal L}^{(k)}(t)+e^{-\gamma_{0}t/2}\frac{\sigma^{(0)}_{-}}{2}\prod^{N}_{k=1}
D_{\cal L}^{(k)\dagger}(t)\right)
\nonumber\\
&+e^{-\gamma_{0}t/2}\frac{\sigma^{(0)}_{+}}{2}\sum^{N}_{k=1}D_{\cal L}^{(1)}(t)\cdots\left(\uline{-i\frac{J_{k}}{2}d^{(k)}_{z}(t)\frac{\sigma^{(k)}_{0}}{2}-\left(i\frac{J_{k}}{2}d^{(k)}_{0}(t)+\gamma_{k}d^{(k)}_{z}(t)\right)\frac{\sigma^{(k)}_{z}}{2}}\right)\cdots D_{\cal L}^{(N)}(t)
\nonumber\\
&+e^{-\gamma_{0}t/2}\frac{\sigma^{(0)}_{-}}{2}\sum^{N}_{k=1}D_{\cal L}^{(1)\dagger}(t)\cdots\left(\uuline{i\frac{J_{k}}{2}d^{(k)\dagger}_{z}(t)\frac{\sigma^{(k)}_{0}}{2}+\left(i\frac{J_{k}}{2}d^{(k)\dagger}_{0}(t)-\gamma_{k}d^{(k)\dagger}_{z}(t)\right)\frac{\sigma^{(k)}_{z}}{2}}\right)\cdots D_{\cal L}^{(N)\dagger}(t),
\end{align}
where we use ${\cal G}[\sigma^{(0)}_{0}]=0$ and $[H^{\rm in}_{k,k'},D_{\cal L}^{(k'')}(t)]=0$ for any $k$, $k'$, and $k''$.
Meanwhile, the derivative of the state is given as
\begin{align}
\frac{d \rho_{{\cal L}}(t)}{d t}=&-\frac{\gamma_{0}}{2}\left(e^{-\gamma_{0}t/2}\frac{\sigma^{(0)}_{+}}{2}\prod^{N}_{k=1}
D_{\cal L}^{(k)}(t)+e^{-\gamma_{0}t/2}\frac{\sigma^{(0)}_{-}}{2}\prod^{N}_{k=1}
D_{\cal L}^{(k)\dagger}(t)\right)\nonumber\\
&+e^{-\gamma_{0}t/2}\frac{\sigma^{(0)}_{+}}{2}\sum^{N}_{k=1}D_{\cal L}^{(1)}(t)\cdots\left(\uline{ \dot{d}^{(k)}_{0}(t)\frac{\sigma^{(k)}_{0}}{2}+\dot{d}^{(k)}_{z}(t)\frac{\sigma^{(k)}_{z}}{2}}\right)\cdots D_{\cal L}^{(N)}(t)\nonumber\\
&+e^{-\gamma_{0}t/2}\frac{\sigma^{(0)}_{-}}{2}\sum^{N}_{k=1}D_{\cal L}^{(1)\dagger}(t)\cdots\left(\uuline{\dot{d}^{(k)\dagger}_{0}(t)\frac{\sigma^{(k)}_{0}}{2}+\dot{d}^{(k)\dagger}_{z}(t)\frac{\sigma^{(k)}_{z}}{2}}\right)\cdots D_{\cal L}^{(N)\dagger}(t).
\end{align}
Comparing the underlined parts corresponding to the basis $\frac{\sigma^{(0)}_{+}}{2}\frac{\sigma^{(k)}_{0,z}}{2}$ (or double-underlined parts corresponding to $\frac{\sigma^{(0)}_{-}}{2}\frac{\sigma^{(k)}_{0,z}}{2}$) in the above two equations,
we obtain the dynamical equation for $\{d^{(k)}_{0}(t),d^{(k)}_{z}(0)\}^{N}_{k=1}$ as
\begin{equation}
\frac{d}{d t}
\begin{pmatrix}
d^{(k)}_{0}(t)\\
d^{(k)}_{z}(t)
\end{pmatrix}
=\frac1 2
\begin{pmatrix}
0&-i J_{k}\\
-i J_{k}&-2 \gamma_{k}
\end{pmatrix}
\begin{pmatrix}
d^{(k)}_{0}(t)\\
d^{(k)}_{z}(t)
\end{pmatrix}
,~\forall k=1,\cdots,N.
\end{equation}
\end{widetext}
Each $k$ component can be solved independently of $k'(\neq k)$ components.
This is easily solved, and its solution is
\begin{align}
d^{(k)}_{0}(t)&=e^{-\gamma_{k} t/2}\bigg(\frac{\gamma_{k}}{\Omega_{k}}
\sin(\Omega_{k}t/2)+\cos(\Omega_{k}t/2)\bigg),
\nonumber \\
d^{(k)}_{z}(t)&=e^{-\gamma_{k} t/2}\bigg( \frac{- i J_{k}}{\Omega_{k}}\sin(\Omega_{k}t/2)\bigg),
\end{align}
where $\Omega_{k}=\sqrt{J^{2}_{k}-\gamma_{k}^{2}}$.
See Ref.~\cite{Kukita_2020} for further details of the derivation of the above solution.
Taking appropriate parameters allows us to reproduce the solutions (\ref{eq:solution0}), (\ref{eq:solution0noise}), (\ref{eq:solution1}), and (\ref{eq:solution1noise}).
For example, when taking $J_{k}=J$ ($\forall k=1,\cdots,N$), $\gamma_{0}=\gamma_{\1}$, $\gamma_{k}=\gamma_{\2}$ ($\forall k=1,\cdots,N$), and $J_{k,k'}=0$ ($\forall k,k'$), we reproduce Eqs.~ (\ref{eq:solution0}) and (\ref{eq:solution0noise}).
When we decompose $N$ qubits into System II and III, and take $J_{k,k'}=0$ for $k,k'$ in the same system, Eqs.~(\ref{eq:solution1}) and (\ref{eq:solution1noise}) are obtained.

\bibliography{relaxation}

\end{document}